\documentclass[titlepage,12pt]{article}
\usepackage{amssymb,epsfig,pslatex}
\usepackage{cite}  
\usepackage{float}       
\usepackage{wrapfig}  
\usepackage{hhline} 
\usepackage{dcolumn} 
\restylefloat{figure} 

\makeatletter
\def\@sect#1#2#3#4#5#6[#7]#8{\ifnum #2>\c@secnumdepth
  \def\@svsec{}\else
  \refstepcounter{#1}\edef\@svsec{\csname the#1\endcsname.\hskip0.5em}\fi
  \@tempskipa #5\relax
  \ifdim \@tempskipa>\z@
    \begingroup
      #6\relax
      \@hangfrom{\hskip #3\relax\@svsec}{\interlinepenalty \@M #8\par}%
    \endgroup
    \csname #1mark\endcsname{#7}\addcontentsline
      {toc}{#1}{\ifnum #2>\c@secnumdepth \else
        \protect\numberline{\csname the#1\endcsname}\fi #7}%
  \else
    \def\@svsechd{#6\hskip #3\@svsec #8\csname #1mark\endcsname
      {#7}\addcontentsline{toc}{#1}{\ifnum #2>\c@secnumdepth \else
        \protect\numberline{\csname the#1\endcsname}\fi #7}}%
  \fi \@xsect{#5}}
\@addtoreset{equation}{section}
\makeatother
\renewcommand\thesection{\arabic{section}}
\renewcommand\theequation{\ifnum \value{section}>0
 \thesection.\arabic{equation}%
\else
\arabic{equation}%
\fi}
\def\shat{\hat{s}}

\def\sp{{\mathbf S}_t}
\newcommand{\sm}{{\bf S}_{\bar t}}
\newcommand{\kh}{{{\bf \hat k}}}

\newcommand{\be}{\begin{equation}}
\newcommand{\ee}{\end{equation}}

\newcommand{\bea}{\begin{eqnarray}}
\newcommand{\eea}{\end{eqnarray}}
\newcommand{\bfig}{\begin{figure}}
\newcommand{\efig}{\end{figure}}
\newcommand{\bc}{\begin{center}}
\newcommand{\ec}{\end{center}}

\newcommand{\mtt}{M_{t \bar t}}

\renewcommand{\thefootnote}{\small\fnsymbol{footnote}}

\begin{document}
\begin{titlepage}
  \begin{flushright}
    PITHA 06/07  \\
    SDU-HEP200606
  \end{flushright}        
\vspace{0.01cm}
\begin{center}
{\LARGE {\bf Weak interaction corrections
to hadronic top quark pair production}}  \\
\vspace{2cm}
{\large{\bf Werner Bernreuther\,$^{a,}$\footnote{Email:
{\tt breuther@physik.rwth-aachen.de}},
Michael F\"ucker \,$^{a,}$\footnote{Email:
{\tt fuecker@physik.rwth-aachen.de}},
Zong-Guo Si\,$^{b,}$\footnote{Email: {\tt zgsi@sdu.edu.cn}} 
}}
\par\vspace{1cm}
$^a$Institut f\"ur Theoretische Physik, RWTH Aachen, 52056 Aachen, Germany\\
$^b$Department of Physics, Shandong University, Jinan, Shandong
250100, China\\
\par\vspace{1cm}
{\bf Abstract}\\
\parbox[t]{\textwidth}
{In this paper we determine the weak-interaction corrections of
order $\alpha_s^2\alpha$ to hadronic 
top-quark pair production. First we 
compute the one-loop  weak corrections to $t \bar t$
production due to gluon fusion and the
order  $\alpha_s^2\alpha$ corrections
to  $t \bar t$ production due to (anti)quark gluon scattering
in the Standard Model. 
With our previous result
\cite{Bernreuther:2005is} this yields the complete 
corrections of order $\alpha_s^2\alpha$ to  $gg$, $q {\bar q}$, $q g$,
and ${\bar q} g$
induced hadronic $t \bar t$ production with $t$ and $\bar t$
polarizations  and spin-correlations fully taken into account. 
For the Tevatron and the LHC we determine  the weak
contributions  to the transverse top-momentum and to the 
$t \bar t$ invariant mass distributions.  At the LHC these corrections can be of the
order of 10 percent compared with the leading-order results, 
for large $p_T$ and $\mtt$, respectively. Apart from
parity-even $t \bar t$ spin correlations we analyze also
parity-violating double and single spin asymmetries, and show how
they are related if CP invariance holds. For $t$ (and $\bar t$) quarks
which decay semileptonically, we compute a resulting charged-lepton
forward-backward asymmetry $A_{PV}$ with
respect to the $t$ ($\bar t$) direction, which is of the order of
one percent at the LHC for suitable invariant-mass cuts.}
\end{center}
\vspace*{2cm}

PACS number(s): 12.15.Lk, 12.38.Bx, 13.88.+e, 14.65.Ha
\end{titlepage}
%
%
\setcounter{footnote}{0}
\renewcommand{\thefootnote}{\arabic{footnote}}
\setcounter{page}{1}
\section{Introduction}
Even after a decade of experimental research
at the Tevatron, the top quark is still 
a relatively unexplored particle as compared to the other quarks and
leptons. The situation will change once the LHC will operate
with planned luminosity,
as it is expected that the large event rates will allow for
precise investigations of these quarks. Full 
exploration of  the data requires also
 precise theoretical predictions for  top quark production and decay,
 especially within the Standard Model (SM).

As far as hadronic top quark pair production is concerned, 
predictions for unpolarized $t \bar t$
production have long been known at  next-to-leading order (NLO) QCD 
\cite{Nason:1987xz,Nason:1989zy,Beenakker:1988bq,Beenakker:1990ma,Mangano:1991jk,Frixione:1995fj}, 
and these NLO results were
refined by resummation of soft gluon and threshold logarithms
\cite{Bonciani:1998vc,Kidonakis:2001nj,Cacciari:2003fi,Kidonakis:2004hr,Banfi:2004xa}.
Moreover, $t \bar t$ production and decay including the full spin
degrees of freedom of the intermediate $t$ and $\bar t$ resonances 
were determined to NLO QCD some time ago 
\cite{Bernreuther:2000yn,Bernreuther:2001bx,Bernreuther:2001rq,Bernreuther:2004jv}.
Top quark spin effects can be reliably predicted in view of the extremely short
lifetime of these quarks, that is, the short-distance nature of their
interactions, and are expected to play an important role in refined data analyses.

A complete NLO analysis of $t \bar t$ production within  the SM should include
also the electroweak interactions.  While 
they are not relevant for the production cross section $\sigma_{t
  \bar t}$ at the Tevatron and at the LHC (see Section 3 below),
they may be important for distributions at large transverse top
momentum or large $t \bar t$ invariant mass due to large Sudakov
logarithms\footnote{See, e.g., \cite{Melles:2001ye,Denner:2000jv,Denner:2006jr}
for reviews and references concerning electroweak Sudakov logarithms.}. Moreover, the weak interactions induce
small parity-violating effects, and for full exploration and interpretation
of future data it is important to obtain definite SM
predictions also for these effects.

Weak interaction corrections to  hadronic $t \bar t$ production were
studied  so far in a number of papers. The order
$\alpha_s^2\alpha$ weak-QCD
corrections to $q {\bar q} \to t \bar t$ and $gg \to t \bar t$ of 
order $\alpha_s^2\alpha$ were analyzed  in \cite{Beenakker:1993yr} (see also
\cite{Kao:1997bs}). Full determinations  of these  corrections to
$q \bar q \to t \bar t (g)$, including the
infrared-divergent box contributions and the corresponding real gluon
radiation were made in \cite{Bernreuther:2005is,Kuhn:2005it}. Recently the 
order $\alpha_s^2\alpha$ corrections to $gg \to t \bar t$ including
the quark triangle diagrams $gg \to Z \to t \bar t$ were investigated
in \cite{Moretti:2006nf}. (Cf. \cite{Maina:2003is,Moretti:2006ea} and 
references therein  for  weak
corrections to other four-parton processes.)
In \cite{Kao:1994rn,Kao:1997bs,Kao:1999kj,Bernreuther:2005is,Beccaria:2005jd} 
parity violation in $t \bar t$ production was
analyzed within the SM. Investigations of non-SM effects include
refs. \cite{Hollik:1997hm,Li:1997gh,Kao:1999kj,Beccaria:2004sx}.

In this paper we present results based on our determination
of the complete  weak interaction
corrections of order $\alpha_s^2\alpha$ to  $gg \to t \bar t$,
to $q {\bar q} \to t \bar t (g)$, and for completeness also
for $g q ({\bar q}) \to t {\bar t} q ({\bar q})$,
   with $t$ and $\bar t$
polarizations  and spin-correlations fully taken into account. 
For $t \bar t$ production at the Tevatron and the LHC we determine the weak
contributions  to a number of top spin-(in)dependent distributions. 
As far as parity-violating effects are concerned,
we derive, for  arbitrary  $t$ and $\bar t$ spin bases, 
a  relation  between  a parity-violating double
spin-asymmetry and corresponding single spin asymmetries. 
For the helicity basis this implies that the parity-violating double
spin-asymmetry is equal to the  
corresponding single spin asymmetry if CP invariance holds in  hadronic $t \bar t$ production.
 Taking into account all
contributions to  order  $\alpha_s^2\alpha$ we compute this 
spin asymmetry for the Tevatron and the 
LHC as a function of the $t \bar t$ invariant mass.
This observable may serve as a useful  tool in
exploring the dynamics of hadronic $t \bar t$ production. For some of the
observables considered in this paper the weak corrections were
analyzed before in the literature; however, these analyses did not
take into account the complete order $\alpha_s^2\alpha$ corrections. 
We compare with these results where possible.

The paper is organized as follows. In Section 2 we present our results
for the  order $\alpha_s^2\alpha$ weak corrections to $gg \to t \bar
t$,  both for the partonic cross section and for $t$, $\bar t$
polarization and $t \bar t$ spin-correlation observables. We discuss
the size of the weak corrections versus leading order (LO) QCD results, and make an explicit
comparison with NLO QCD in the case of the partonic cross section. 
For the reactions $g q ({\bar q}) \to t {\bar t} q ({\bar q})$
we calculate the  order $\alpha_s^2\alpha$ weak corrections to the partonic
cross sections and to $t$ and $\bar t$ spin observables.
Then 
we derive, for the partonic
collisions that initiate $t \bar t$ production, 
several relations involving top-spin observables.
In particular, we show that
the  parity-violating single and double spin-asymmetries are not
independent observables. 
In this and in the next section we comment also briefly on a
CP-violating asymmetry. In Section 3 we give our
results at the level of hadronic collisions. We determine the 
order $\alpha_s^2\alpha$ weak corrections to the $t \bar t$ cross
section at the Tevatron and at the LHC. We compute the weak
interaction corrections to the transverse top-momentum and the
$t \bar t$ invariant mass distribution, and to two parity-invariant 
double spin asymmetries. Moreover, we determine the above-mentioned
 parity-violating double spin asymmetry in the helicity basis, which
is equal to the corresponding single $t$ spin asymmetry, as a 
function of $M_{t \bar t}$ and determine the resulting
charged-lepton forward-backward asymmetry for semileptonic
$t$ quark decays. We conclude in Section 4.
 \section{Parton level results}
At the Tevatron and at the LHC top quark pairs
are produced predominantly by the strong interactions. Theoretical
predictions for the  subprocesses $i \to t {\bar t}+X$, $(i=q {\bar
q}, gg, gq, g {\bar q})$  are known to order $\alpha_s^3$. 
The leading corrections to these parton processes involving
electroweak interactions are, for $i=q {\bar q}$,
the order  $\alpha^2$  Born contributions  (from 
$q \bar q \to \gamma, Z \to t \bar t$) and,  for $i=q {\bar q}$ and
$gg$, the mixed QCD electroweak corrections of order  $\alpha_s^2 \alpha$.
Due to color conservation there are no corrections of order
$\alpha_s \alpha$. 

\begin{figure}[H]
\begin{center}
\epsfig{file=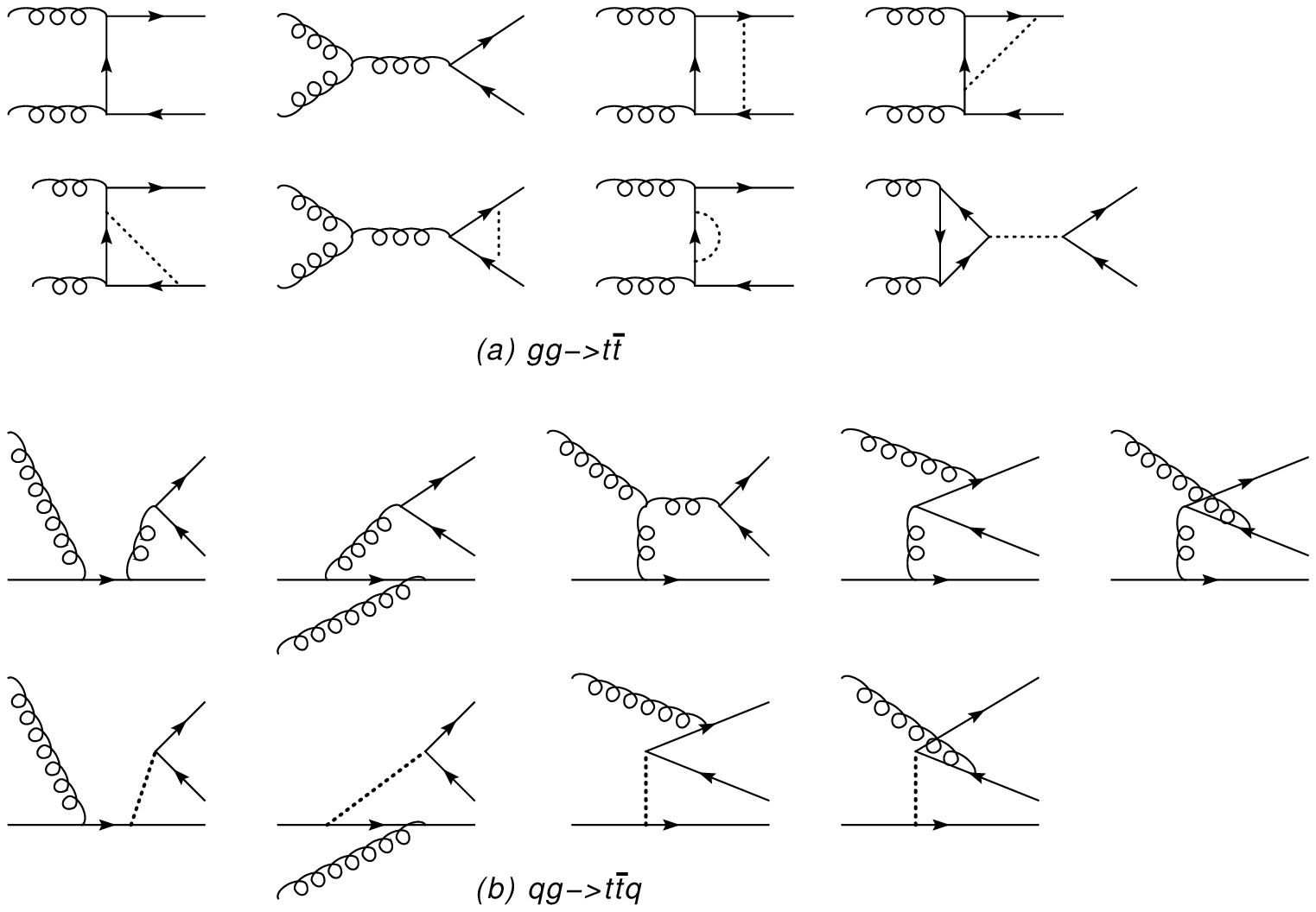, width=12cm, height=10cm}
\end{center}
\caption
{ (a) Lowest order QCD diagrams and 1-loop weak corrections
to $gg \to t {\bar t}$. Crossed diagrams are not drawn. The dotted
line in the box diagram and in the vertex and self-energy corrections
represents $W,\, Z$ bosons, the corresponding Goldstone bosons, and the
Higgs boson $H$. The fermion triangle in the last diagram represents a
$t$ and $b$ quark loop, followed by $s$ channel exchange of the $Z$ boson,
the associated Goldstone boson, and the Higgs boson.  
(b) Tree-level diagrams for $q g \to t {\bar t} q$. Upper row: QCD
diagrams; lower row: mixed electroweak-QCD contributions. The dotted
line represents a photon or a $Z$ boson.
} \label{fig:diagramr}
\end{figure}

In \cite{Bernreuther:2005is} we have determined the weak corrections 
of order $\alpha_s^2 \alpha$ for $q \bar q$ initiated top-pair
production, which involve the reactions $q {\bar q} \to t \bar t$
and $q {\bar q} \to t {\bar t} g$; see also \cite{Kuhn:2005it}.
Here we consider gluon-gluon fusion,
\begin{equation}
g(p_1)+g(p_2)\rightarrow t(k_1, s_t)+\bar{t}(k_2, s_{\bar t})\, ,
\label{ggs12}
\end{equation}
and we present in this section 
our  results for the weak interaction effects
on the cross section and on several single and double spin observables
of this parton reaction.
In (\ref{ggs12}) the parton
momenta  are denoted by $p_1$, $p_2$, $k_1$, and $k_2$,  and
the vectors $s_t$,  $s_{\bar t}$,
with  $s^2_t = s^2_{\bar t} = -1$ and $ k_1\cdot s_t = 
k_2\cdot  s_{\bar t} = 0$ describe the spin of the top and antitop quark.
The leading  correction  involving
electroweak interactions\footnote{Although we
consider in this paper purely weak corrections, we
parameterize our results for convenience in terms of the
QED coupling $\alpha=\alpha_W \sin^2\theta_W$.} 
to the differential cross section of (\ref{ggs12}) is of the form
\be
\alpha_s^2 \alpha \; \delta {\cal
  M}_{W}(p,k,s_t, s_{\bar t}) \, .
\label{convir}
\ee
We are interested here only in purely weak, in particular 
in parity-violating
effects. Therefore we take into account only the mixed QCD and weak contributions to  $\delta
{\cal M}_{W}$ 
in the following. The photonic  contributions 
form a gauge invariant set and can be straightforwardly
obtained separately. 
The contributions to $\delta {\cal M}_W$ are the $gg \to t \bar t$
QCD Born diagrams interfering with
the 1-loop diagrams involving the weak gauge boson, Goldstone
boson (we work in the 't Hooft Feynman gauge), and Higgs boson
exchanges which yield top quark self-energy, vertex, box-diagram,
and, via quark triangle diagrams,  s-channel $Z-$ and Higgs-boson  
contributions \cite{Beenakker:1993yr}. 
The diagrams are shown in Fig.~\ref{fig:diagramr}a.
The ultraviolet
divergences  in the self-energy and vertex corrections are removed using the
on-shell scheme  \cite{Beenakker:1993yr}
for defining the wave function renormalizations
of $t_L$ and $t_R$ and the top quark mass $m_t$. All the 1-loop purely
weak contributions to (\ref{ggs12}) are infrared-finite.

We take into account also the 1-loop amplitudes 
$gg \to Z, \chi^0 \to t \bar t$, where $Z$ denotes an off-shell $Z$ boson and  $\chi^0$ is
the corresponding neutral Goldstone boson. The $gg \to Z$ vertex is induced
by the  flavor non-singlet neutral axial vector
current  $J_{5\mu}^{NS} =  \sum_{i=1}^3
 {\bar \psi_i}\gamma_{\mu}\gamma_5 \tau_3 \psi_i$,
where $\psi_i = (u,d)_i$ is the i-th  generation quark isodoublet and $\tau_3$ is
the 3rd Pauli matrix. Because of the large $t, b$ quark-mass
splitting, only the contribution of the third quark generation
matters. In this paper we take all quarks but the top quark to be massless.
The $gg \to \chi^0$ vertex is generated by
the corresponding pseudoscalar current  $J_{5}^{NS}$. The 
contribution of $gg \to Z, \chi^0 \to t \bar t$ to $\delta {\cal M}_W$
-- which we shall denote by ``non-singlet neutral current contribution''
in the following --
was apparently not considered in \cite{Beenakker:1993yr}, but was
taken into account in the recent paper \cite{Moretti:2006nf}.

We have determined  (\ref{convir}) analytically for
arbritrary $t$ and $\bar t$ spin states. From this expression one
can extract the weak interaction corrections to
the $gg \to t \bar t$  spin density matrix. This matrix, when  
combined with 
the decay density matrices
describing
semi- and non-leptonic $t$ and $\bar t$ decay  yields
predictions at the level of the $t$ and/or  $\bar t$ decay products
of the order $\alpha_s^2 \alpha$ weak interaction effects in top-pair
production.
Likewise one may proceed with  the weak corrections to 
$q {\bar q} \to t \bar t (g)$ \cite{Bernreuther:2005is}.

For the sake of brevity we do not give here
the expression for  $\delta {\cal M}_W(p,k,s_t, s_{\bar t})$,
but present results for the weak corrections
to the partonic cross section  and to several single and double spin
asymmetries, which we believe are of interest to phenomenology. 
The inclusive, spin-summed cross section 
for  (\ref{ggs12})  may be written,
to NLO in the SM gauge couplings, in the form 
\begin{equation}
{\sigma}_{gg} \; = \; {\sigma}_{gg}^{(0)} + 
\delta {\sigma}_{gg}^{(1)} + \delta {\sigma}_{gg}^{W} \; ,
\label{mixed}
\end{equation}
where the first and  second term are  the LO (order $\alpha_s^2)$ and
NLO (order $\alpha_s^3)$ QCD contributions 
\cite{Nason:1987xz,Nason:1989zy,Beenakker:1988bq,Beenakker:1990ma,Bernreuther:2001bx}, and the third
term  denotes  the weak corrections  described above. We parameterize this term
as follows:
\begin{equation}
\delta {\sigma}_{gg}^W(\shat,m^2_t) \; = \; 
\frac{4\pi\alpha_s^2 \alpha}{m^2_t}  \, f^{(1\,W)}_{gg}(\eta) \, ,
\label{eq:xsection}
\end{equation}
where $\eta ={\shat}/{4m^2_t} -1$, with ${\shat}$ being the
gluon-gluon center-of-mass (c.m.) energy  squared.  We have  numerically evaluated the 
scaling function  $f^{(1\,W)}_{gg}(\eta)$ --
and those defined below -- and parameterized them in terms of fits
 which allow for a quick  use in applications. 
In the following we use  $m_Z=91.188$ GeV,
$sin^2\theta_W =0.231$, and  $m_t = 172.7$ GeV
 \cite{unknown:2004qh,unknown:2005cc}.
As already mentioned,  all  quarks but the top quark are taken to be massless. 
Moreover, we use two values of the Higgs boson mass, $m_H=120$ GeV
and $m_H=200$ GeV, which correspond approximately to the present
experimental lower
and upper bound on $m_H$. Moreover,
$\alpha_s(2m_t)=0.1$ and $\alpha(2m_t) = 1/126.3$ were
chosen  in the results given below. In the $s$-channel Higgs exchange
diagram we take into account the finite width of the Higgs boson;
however, for the chosen range of $m_H$ this is numerically insignificant.

\begin{figure}[H]
\begin{center}
\epsfig{file=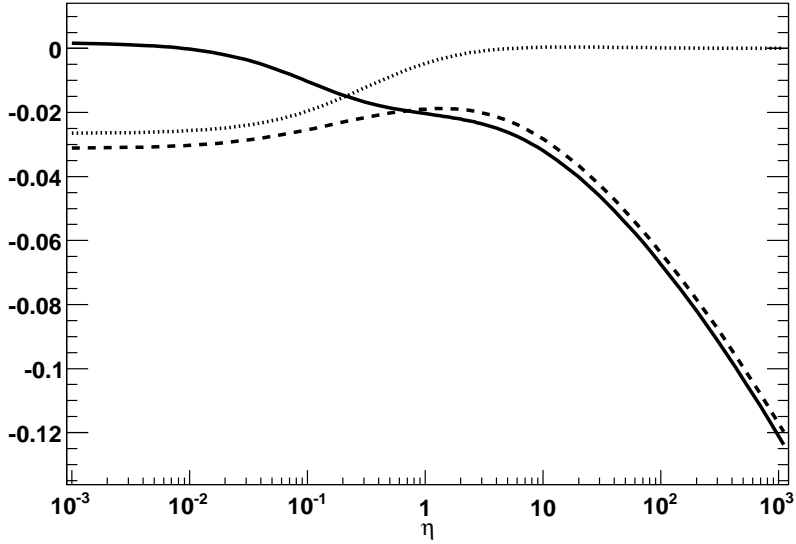, width=12cm ,height=8cm}
\end{center}
\caption{
Ratio $r_W^{(0)}$ of the order $\alpha \alpha_s^2$ corrections and the Born
cross section for $gg \to t {\bar t}$ as a function of $\eta$,
for two Higgs masses,  $m_H=120$ GeV (solid) and $m_H=200$ GeV (dashed).
The dotted  curve shows the non-singlet neutral current contribution
to $r_W^{(0)}$.
}\label{fig:rweakgg}
\end{figure}

In Fig.~\ref{fig:rweakgg} we have plotted the ratio
\begin{equation}
r_W^{(0)} \; = \; 
\frac{ \delta {\sigma}_{gg}^{W}  }{{\sigma}_{gg}^{(0)} }  
\label{eq:rw}
\end{equation}
as a function of $\eta$ for the two Higgs masses given above.
This figure  shows that the
non-singlet neutral current contribution is relevant, as compared to
the other weak corrections,  in the vicinity of the $t \bar t$
threshold up to $\eta \sim 1$. The weak interaction corrections to
$\sigma_{gg}$ are essentially negative for all Higgs-boson
masses above the present experimental lower bound -- except for
$m_H \leq 120 GeV$ very close to threshold.
The weak corrections (\ref{eq:xsection}) to the
cross section  of the $gg$ subprocess and $r_W^{(0)}$
have recently been computed also 
by \cite{KSU}. We have compared our results and find excellent
numerical agreement. 
Moreover, we have evaluated our results for
(\ref{eq:xsection}), excluding the non-singlet neutral current
contribution, with the parameter values 
chosen in  \cite{Beenakker:1993yr} and compared with the results
given in Figs.~11 - 16 of that paper, with which we also agree.

From Fig.~\ref{fig:rweakgg} one might conclude that for large
$\sqrt{\hat s}$ the weak corrections to $\sigma_{gg}$ grow as compared
with the QCD cross section. However, this is deceptive:  the NLO
QCD corrections must be taken into account for a
realistic assessment of the high energy behavior. Real gluon radiation, $gg 
\to t {\bar t} g$, involves $t$- and $u$-channel
gluon exchange diagrams, which are dominant at high energies, while
such exchanges of massless spin-one particles 
are absent at lowest order QCD and for the weak
corrections of order $\alpha_s^2 \alpha$. This causes the 
NLO QCD corrections to $\sigma_{gg}$ to approach a  constant for large $\hat s$
\cite{Nason:1987xz}
while the Born cross section falls off. 
Thus the NLO QCD corrections to the $gg$ initiated $t \bar t$ 
production show a 
high-energy behaviour\footnote{This applies also to $gq ({\bar q})$
  initiated $t \bar t$ production \cite{Nason:1987xz}.} which is strikingly different from  the QCD
corrections to  $q \bar q$ induced top-pair production, which fall off
for large $\sqrt{\hat s}$.  Moreover, we recall that the NLO QCD
corrections to $q \bar q$ and $gg$ initiated cross sections are large
in the vicinity of the $t \bar t$ threshold due to the exchange of
Coulomb gluons. 

\begin{figure}[H]
\begin{center}
\hspace*{-1.5cm} \epsfig{file=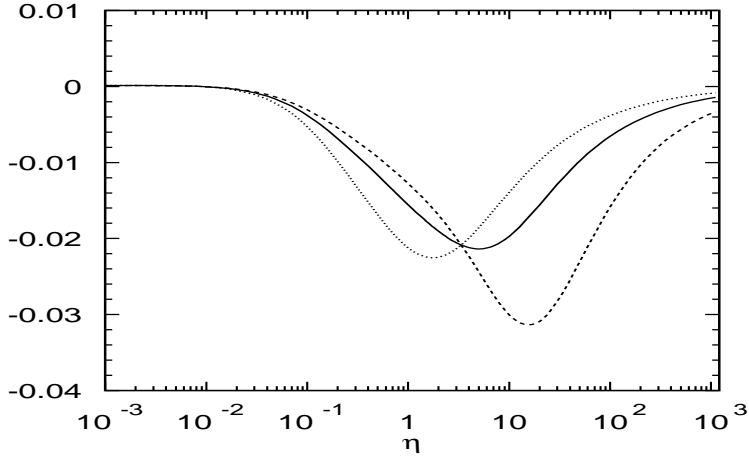, width=11.3cm, height=7cm}
\end{center}
\caption{Ratio 
$r_W^{(1)}$ of the order $\alpha \alpha_s^2$ corrections (for
$m_H=120$ GeV)
and the NLO QCD cross section for $gg \to t {\bar t}$ (taken
  from \cite{Bernreuther:2001bx,Bernreuther:2004jv}), evaluated 
for $\mu = m_t/2$ (dotted), $\mu = m_t$ (solid), and $\mu = 2m_t$ (dashed).
}\label{fig:rw1gg}
\end{figure}

Fig.~\ref {fig:rw1gg} exhibits the ratio 
\begin{equation}
r_W^{(1)} \; = \; 
\frac{ \delta {\sigma}_{gg}^{W}  }{{\sigma}_{gg}^{(0)} + \delta {\sigma}_{gg}^{(1)}  }  
\label{eq:rs}
\end{equation}
as a function of $\eta$ for $m_H=120$ GeV and
three values of the renormalization scale
$\mu$, which is  put equal to the factorization scale. 
In this Figure the coupling  $\alpha_s(\mu)$ has been evaluated  according to two-loop renormalization
group evolution. Fig.~\ref{fig:rw1gg} shows that the weak corrections
to $\sigma_{gg}$, which are negative,  do not exceed $\sim 3 \%$ in magnitude. The size
and location of the maximum of $|r_W^{(1)}|$ depends on the scale $\mu$.
Eventually, the significance of the weak corrections must be
investigated at the level of hadronic collisions.

\begin{figure}[H]
\begin{center}
\epsfig{file=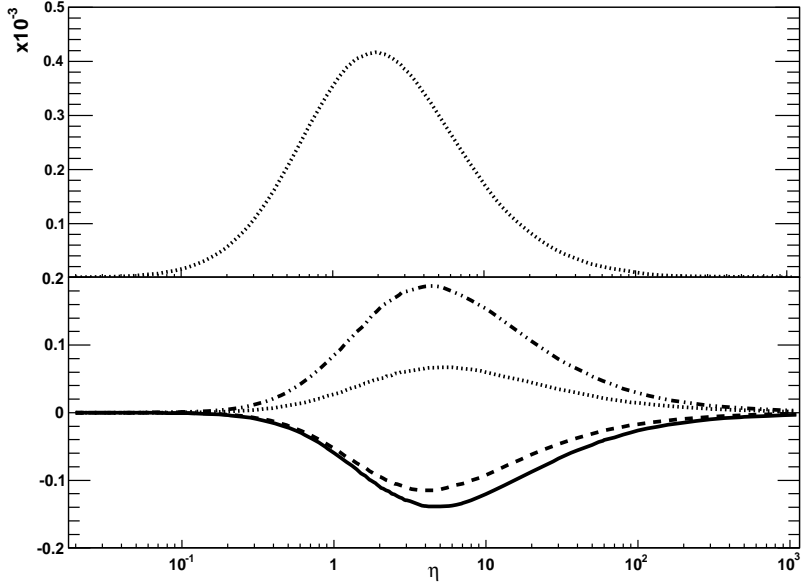, width=12cm,height=8cm }
\end{center}
\caption{Upper frame: Scaling function $f^{(1W)}_{ug}$ defined in
(\ref{eq:qgsection}) for u-type quarks.  For d-type quarks, $f^{(1W)}_{dg}=
-f^{(1W)}_{ug}$. Lower frame:
Scaling functions  $h^{(1\, W,hel)}_{ug}$ (dotted),
 $h^{(1\, W,hel)}_{dg}$  (solid),
$h^{(1\, W,hel)}_{{\bar u} g}$  (dashed), and
$h^{(1\, W,hel)}_{{\bar d} g}$ (dash-dotted) that determine
the expectation value   (\ref{eq:qgbarspin}) for the helicity axis.
 }\label{fig:gqfh1}
\end{figure}

For completeness we have
determined the order $\alpha_s^2\alpha$ corrections to the
partonic cross sections
also for the reactions $g q ({\bar q}) \to t {\bar t} q ({\bar
  q})$. These corrections arise from 
the interference of the QCD and the mixed electroweak-QCD
diagrams shown in  Fig.~\ref{fig:diagramr}b. Notice that the diagram
involving the three-gluon vertex does not interfere with the mixed diagrams due to
color mismatch. Writing 
${\sigma}_{qg}  =  {\sigma}_{qg}^{(1)} + \delta
{\sigma}_{qg}^{W}$, where  ${\sigma}_{qg}^{(1)}$ is the order
$\alpha_s^3$
 Born cross section, we parameterize  $\delta {\sigma}_{qg}^{W}$
in analogy to (\ref{eq:xsection}):
\begin{equation}
\delta {\sigma}_{qg}^W(\shat,m^2_t) \; = \; 
\frac{4\pi\alpha_s^2 \alpha}{m^2_t}  \, f^{(1\,W)}_{qg}(\eta) \, .
\label{eq:qgsection}
\end{equation}
From crossing symmetry we have $f^{(1\,W)}_{{\bar q}g}(\eta)=
f^{(1\,W)}_{qg}(\eta)$.
The scaling function for u-type quarks,  $f^{(1\,W)}_{qg}(\eta)$
is shown in Fig.~\ref{fig:gqfh1}, upper frame.
For d-type quarks $f^{(1\,W)}_{dg}(\eta)= -
f^{(1\,W)}_{ug}(\eta)$ holds. This is due to the fact that in the
interference terms of the weak interaction diagrams, involving the
$\gamma t {\bar t}$ and $Zt{\bar t}$ vertices, and the QCD diagrams
the terms that are generated by the vector currents vanish due to Furry's
theorem, and $f^{(1\,W)}_{qg}$ is proportional to $a_q a_t$, where
$a_q$ is the neutral-current axial-vector coupling. For the Tevatron
and
the LHC the corrections (\ref{eq:qgsection}) are small as
compared to  ${\sigma}_{qg}^{(1)}$, which in turn  makes
only a
small contribution to the $t \bar t$ cross section, as compared with
$gg$ and $q \bar q$ initiated production. For the hadronic
cross section to be discussed in the next section,
we will therefore take into account only these initial parton states.
 
Next we consider observables that involve the $t$ and/or $\bar t$
spin. Denoting the top spin operator by
$\sp$ and its projection onto an arbitrary unit vector ${\bf\hat a}$
by $\sp \cdot {\bf\hat a}$ we can express 
its unnormalized  partonic expectation
value, which we denote by double brackets, in terms of the
difference between the ``spin up'' and ``spin down'' cross sections:
\begin{equation}
  2 \langle  \langle \sp \cdot {\bf\hat a}  \rangle  {\rangle}_i
  = \sigma_i(\uparrow ) - \sigma_i(\downarrow) \,  ,
\label{sspas}
\end{equation}
where $\langle  \langle {\cal O}  \rangle  {\rangle}_i \equiv
\int d\sigma_i {\cal O}$.
Here $i$ denotes one of the
partonic initial states that produce $t \bar t$,
 and the arrows refer to the projection of the top-quark spin
onto ${\bf{\hat a}}$. An analogous formula  holds for the antitop
quark. It is these expressions  that enter the 
corresponding predictions at the level of hadronic  collisions.

There are two types
of single spin asymmetries (\ref{sspas}):
parity-even, T-odd asymmetries\footnote{T-even/odd
refers to the behavior with respect to a naive T transformation,
i.e., reversal of momenta and spins.}, where the spin projection is onto 
an axial vector, and
parity-odd, T-even  ones where ${\bf\hat a}$ is a polar vector. 
The asymmetry associated with
the projection of $\sp$ onto the normal of the $q, t$ scattering
plane belongs to the first class.
It is induced by the parity-even absorptive part of $\delta {\cal M}_W$,
but also by the absorptive part of the NLO QCD amplitude.  The
QCD-induced $t$ and $\bar t$ polarization normal to the scattering
plane is of the order of a few percent 
\cite{Bernreuther:1995cx,Dharmaratna:xd}. The weak
contribution is even smaller; therefore we do not display it here.  

\begin{figure}[H]
\begin{center}
\epsfig{file=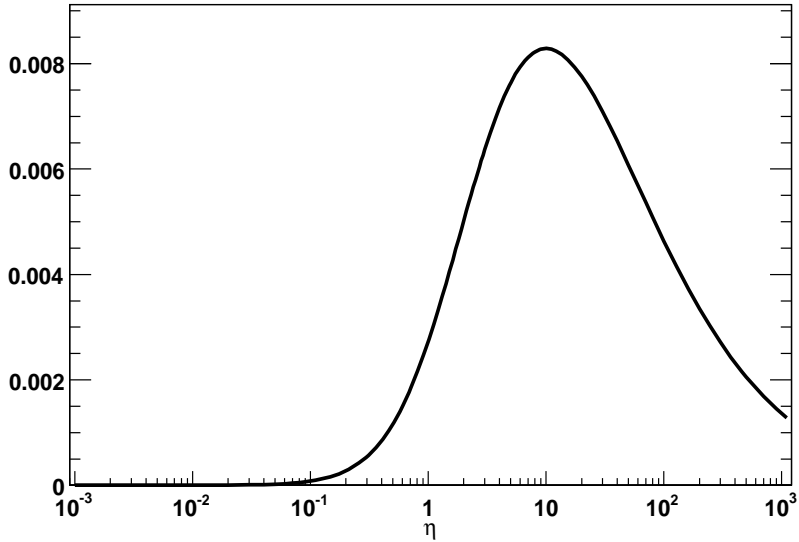, width=12cm,height=8cm }
\end{center}
\caption{Scaling function  $h^{(1\, W,hel)}_{gg}$   that determines
the expectation value   (\ref{eq:sinspin}) for the helicity axis.
 }\label{fig:ggheli}
\end{figure}

The P-odd, T-even  single spin asymmetries correspond to a
polarization
 of the $t$ (and $\bar t$) quarks along a polar vector, in particular along 
a direction in the scattering plane. Needless to say, these asymmetries
cannot be generated within QCD; the SM contribution  results from the parity-violating part of
 $\delta{\cal M}_W$. 
Popular choices are  top-spin projections onto the beam axis \cite{Bernreuther:2004jv}
and the  off-diagonal axis  ${\bf\hat d}_{{\rm off}}$ \cite{Mahlon:1997uc},
which are relevant
for the Tevatron, and  onto the helicity axes, which are relevant for the LHC.
These  axes must be defined in a collinear safe reference frame, and
a convenient one with this property is the $t \bar
t$ zero-momentum frame (ZMF) \cite{Bernreuther:2004jv}.
With respect to this frame we define  
\begin{eqnarray}
     {\bf\hat a} = {\bf\hat b} = {\bf\hat p}, &&\mbox{(beam\
      basis)}
\label{beambasis},\\
        {\bf\hat a} = -  {\bf\hat b} = \kh,&&
    \mbox{(helicity basis)}
\label{helbasis}, 
\end{eqnarray}
where $\kh$ denotes the direction of
flight of the top quark in the  $t\bar{t}$ ZMF
and ${\bf\hat p}$ is the direction of flight of one of the colliding
hadrons in that frame. The direction of the hadron beam can be identified to
a very good approximation with the direction of flight of one of the
initial partons. The unit vector ${\bf\hat b}$ serves as
quantization axis for the $\bar t$ quark spin. In fact, the beam and
off-diagonal axes are useless here, as we have the result (which is
exact for the $gg \to t \bar t$ amplitude):
\begin{equation}
   \langle  \langle \sp \cdot {\bf\hat p}  \rangle  {\rangle}_{gg}
  \, =  \,  \langle  \langle \sp \cdot {\bf\hat d}_{{\rm off}}  \rangle
  {\rangle}_{gg}    \, = \, 0            \,  .
\label{bdoffres}
\end{equation}
Eq.~(\ref{bdoffres}) follows from the properties of the coefficients
of the $gg \to t \bar t$ spin density matrix
dictated by Bose symmetry of the initial $gg$ state, which were 
derived in \cite{Bernreuther:1993hq}.
As to the helicity basis, 
the unnormalized  expectation value of 
${\bf S_t}\cdot {\bf\hat k}$ is again conveniently expressed by a scaling
function:
\begin{equation}
\langle \langle 2 {\sp}\cdot {\bf\hat k} \rangle
{\rangle}_{gg}  =\frac{4\pi \alpha_s^2 \alpha}{m^2_t}
\, h^{(1\,W, hel)}_{gg}(\eta) \, .
\label{eq:sinspin}
\end{equation}
The scaling function $h^{(1\,W, hel)}_{gg}$ is
 shown in Fig.~\ref{fig:ggheli}.  Notice that 
$\langle \langle 2 {\sp}\cdot {\bf\hat a} \rangle \rangle$ does not
depend on $m_H$, as the SM Higgs boson exchange is parity-conserving.
While the non-singlet neutral current diagrams do contribute to 
$\sigma_{gg}$ and to several spin-correlation observables, they have
no effect on $h^{(1\,W, hel)}_{gg}$. This follows from the structure
of the non-singlet neutral-current contribution to
$\delta{\cal M}_W$. 

We have  computed the expectation value of this observable also for 
$q g$ and ${\bar q} g$ initiated $t \bar t$ production. Using a
parameterization analogous to (\ref{eq:sinspin}),
\begin{equation}
\langle \langle 2 {\sp}\cdot {\bf\hat k} \rangle
{\rangle}_{j}  =\frac{4\pi \alpha_s^2 \alpha}{m^2_t}
\, h^{(1\,W, hel)}_{j}(\eta) \, , \quad j= qg, \, {\bar q} g,
\label{eq:qgbarspin}
\end{equation}
the scaling functions  $h^{(1\,W, hel)}_{j}$ are shown in
Fig.~\ref{fig:gqfh1}, lower frame, for $j=ug, {\bar u}g, d g,$ and
${\bar d} g$. Notice that the expectation value 
of ${\sp}\cdot {\bf\hat k}$ is different for $q g$ and ${\bar q} g$
initiated reactions.

Next we analyze top-antitop spin correlations. 
The most interesting set of spin observables besides (\ref{sspas})
seem to be,
as far as SM weak interaction effects are concerned, 
parity-violating double spin asymmetries
defined by the following difference of spin-dependent cross sections:
\begin{equation}
{D}_i(\uparrow\downarrow) \equiv {\sigma}_i({\uparrow\downarrow}) - 
{\sigma}_i(\downarrow \uparrow) \, , \quad i= q{\bar q}, gg , g q, g
{\bar q}, 
\label{dupdown}
\end{equation}
where the first (second) arrow on the right-hand side of
(\ref{dupdown}) refers to the $t$ $({\bar t})$ spin projection onto
a polar vector ${\bf\hat a}$ $({\bf\hat b})$. 
It is obvious that a nonzero ${D}_i(\uparrow\downarrow)$
requires P-violating interactions;  there are, as in the
case of  (\ref{eq:sinspin}), no QCD and
QED contributions to (\ref{dupdown}) to any order in the gauge
couplings. \\
There is a useful relation between these P-violating
double spin asymmetries  and the single-spin observables discussed above.
Using the consequences of rotational invariance for the
 $i \to t {\bar t} \, + X$ spin density matrices 
\cite{Bernreuther:1993hq}, we obtain the following exact result: 
\begin{equation}
2 {D}_i(\uparrow\downarrow) =    \langle \langle 2 {\sp}\cdot {\bf\hat a}
- 2 {\sm}\cdot {\bf\hat b}\rangle {\rangle}_{i}  \, , \quad i= q{\bar
  q}, gg , g q, g {\bar q}.
\label{exaktres}
\end{equation}
That is, the asymmetries (\ref{dupdown}) are completely determined by
the corresponding single $t$ and $\bar t$ spin observables. \\
For the LHC  useful reference axes ${\bf\hat a},$ ${\bf\hat
  b}$ are the helicity axes (\ref{helbasis}). In this case we use the notation
${D}_i(\uparrow\downarrow) = {D}_{RL,i}$, and  (\ref{exaktres})
reads:
\begin{equation}
2 {D}_{RL,i} \, = \, \langle \langle 2 {\sp}\cdot {\bf\hat k}_t
- 2 {\sm}\cdot {\bf\hat k}_{\bar t}\rangle {\rangle}_{i}\, = \,
 \langle \langle (2 {\sp}+ 2\sm)\cdot {\bf\hat k} \rangle
{\rangle}_{i} \: ,
\label{eq:DRLi}
\end{equation}
where $\bf\hat k$ is the $t$ direction in the $t \bar t$ ZMF.
If the interactions that affect $i\to  t {\bar t} \, + X$
are CP-invariant then 
\begin{equation}
{D}_{RL,i} \, = \,\langle \langle 2 {\sp} \cdot {\bf\hat k}_t \rangle
{\rangle}_{i}  \, = \, - \langle \langle 
2 {\sm}\cdot {\bf\hat k}_{\bar t} \rangle {\rangle}_i \, , \quad  i=
q{\bar q}, gg\, ,
\label{cp1rel}
\end{equation}
must hold. Eq.~(\ref{cp1rel}) constitutes a  CP-symmetry test  in 
$t \bar t$ production. In fact, SM CP violation, i.e., the
Kobayashi-Maskawa phase leads to tiny
CP-violating effects (which are induced beyond the 1-loop approximation) in
flavor-diagonal reactions like those considered here. Thus
an experimentally detectable violation
of Eq.~(\ref{cp1rel})  requires non-standard CP-violating interactions
(see below). \\
Let us, for completeness, also discuss the expectation value of the
single spin observable for the reactions $q g, {\bar q} g \to t {\bar
  t} X$. For CP invariant interactions we obtain:
\begin{equation}
 \,\langle \langle 2 {\sp} \cdot {\bf\hat k}_t \rangle
{\rangle}_{q({\bf p}_1) g({\bf p}_2) }  \, = \, - \langle \langle 
2 {\sm}\cdot {\bf\hat k}_{\bar t} \rangle {\rangle}_{{\bar q}(-{\bf
    p}_1) g(-{\bf p}_2) } \, = \, - \langle \langle 
2 {\sm}\cdot {\bf\hat k}_{\bar t} \rangle {\rangle}_{{\bar q}({\bf p}_1) g({\bf p}_2) } \, ,
\label{cpqgrel}
\end{equation}
and an analogous relation holds for
$\langle \langle 2 {\sp} \cdot {\bf\hat k}_t \rangle
{\rangle}_{{\bar q}g}$.
The last equation in (\ref{cpqgrel}) follows from rotational
invariance. These relations and (\ref{exaktres}) imply
\begin{equation}
{D}_{RL,\, qg} \, = \, {D}_{RL,\, {\bar q}g} \, = \, 
\langle \langle {\sp}\cdot {\bf\hat k}_t\rangle {\rangle}_{qg}+
\langle \langle {\sp}\cdot {\bf\hat k}_{ t}\rangle {\rangle}_{{\bar
    q}g} \, .
\label{eq:DRLqgbarq}
\end{equation}

For $i= q{\bar q}, gg$ we parameterize  ${D}_{RL,i}$ as  follows:
\begin{equation}
 {D}_{RL,i} \, = \,  \frac{4\pi \alpha}{m^2_t}[
\alpha \, {\tilde h}^{(0\, W,hel)}_i(\eta) + \alpha_s^2 \, {\tilde
  h}^{(1\, W,hel)}_i(\eta)] \, , 
\label{eq:sinphel}
\end{equation}
where the order $\alpha^2$ term is present only for $i= q \bar q$.
As the SM amplitudes are CP-invariant to NLO in the weak interactions,
Eq.~(\ref{cp1rel}) holds and we obtain the  relations
\begin{equation}
 {\tilde h}^{(0\, W,hel)}_{q \bar q}=   {h}^{(0 \, W,hel)}_{q \bar q}, \quad
{\tilde h}^{(1\,W ,hel)}_i = {h}^{(1\, W,hel)}_{i}, \quad i =q {\bar q}, \, gg \, ,
\label{relPV}
\end{equation}
where the ${h}_{q \bar q}^{(0\, W,hel)},\, {h}_i^{(1\, W,hel)}$ are the scaling functions
of the  single-spin observable (\ref{sspas})
in the helicity basis. For the $q \bar q$
initial state they were given\footnote{The results for the single-spin
  scaling functions $h_{q{\bar q}}$ given in \cite{Bernreuther:2005is}
must be multiplied by a factor 2.}
 in \cite{Bernreuther:2005is}, and for
$gg$ fusion it is shown in Fig.~\ref{fig:ggheli}. \\
The asymmetry ${D}_{RL,i}$, which for $i=gg, q{\bar q}$ is invariant under  a CP transformation,
should not be confused with  the following P- and CP-odd, but
T-even spin asymmetry \cite{Bernreuther:1993df,Schmidt:1992et}:
\begin{equation} 
 \langle \langle (2 {\sp} -  2\sm)\cdot {\bf\hat k} \rangle
{\rangle}_{i} \, = \, 2 {\sigma}_{i,++} - 
2 {\sigma}_{i,--} \, , \quad i= q{\bar q}, gg ,
\label{eq:cpas}
\end{equation}
where the first (second) subscript refers to the $t$ $({\bar t})$
helicity. A non-zero value of (\ref{eq:cpas}) -- or
 equivalently, a violation of
(\ref{cp1rel}) -- requires CP-violating absorptive parts in
the scattering amplitude. These may be generated, for instance, by
non-standard neutral Higgs bosons with both scalar and pseudoscalar
couplings to top quarks
\cite{Bernreuther:1993df,Bernreuther:1993hq,Schmidt:1992et}. \\
For arbitrary reference axes  ${\bf\hat a}$, ${\bf\hat b}$ the
analogue of (\ref{eq:cpas}) reads:
\begin{equation} 
 \langle \langle (2 {\sp}\cdot{\bf\hat a}  +  2\sm \cdot{\bf\hat b}  \rangle
{\rangle}_{i} \, = \, 2 {\sigma}_i(\uparrow \uparrow) - 
2 {\sigma}_i(\downarrow \downarrow)\, .
\label{eq:calgem}
\end{equation}

Finally  we analyze the weak interaction contributions to 
parity- and T-even $t \bar t$ spin-correlation observables,
which are generated already to
lowest order QCD. For the Tevatron
these spin correlations (including NLO corrections) are
largest with respect to the beam and off-diagonal bases, while for the
LHC the helicity basis is a good choice\footnote{For the LHC, a basis
has been  constructed \cite{Uwer:2004vp} which gives a QCD  effect which
is  somewhat larger than using  the helicity correlation.}. In addition, as
was shown in \cite{Bernreuther:2004jv}, a
good measure for the  spin correlation of the $t \bar t$ pair
produced at the LHC is
 the distribution
of  the opening angle between the two particles/jets
from $t$ and $\bar t$ decay that are used as top-spin analyzers.
A non-uniform  distribution is due to the
correlation ${\bf S}_t\cdot {\bf S}_{\bar t}$. Within the SM these spin correlations
 result, for most values of the
parton c.~m. energy squared which are accessible at the
Tevatron and at the LHC,  almost exclusively from the strong
interaction dynamics; the effect of the weak interactions on these observables
turns out to be small. As the precise measurement of these correlations is
expected to be  feasible only at the LHC, we display here results only
for two observables which are useful for data analysis
at this collider.  These are the helicity correlation and the spin-spin projection
mentioned above, which we denote, using the
convention of \cite{Bernreuther:2005is}, by
\begin{equation}
{\cal O}_3 \equiv - 4\,(\kh\cdot\sp)(\kh \cdot\sm),
\label{eq:hbasis}
\end{equation}
\begin{equation}
{\cal O}_4 \equiv 4\,\sp\cdot\sm = 4 \sum_{i=1}^3 ({\bf{\hat e}}_i
\cdot \sp)
  ({\bf{\hat e}}_i\cdot \sm)\, ,
\label{eq:sbasis}
\end{equation}
where $\kh$ denotes as before the $t$ direction
  in the $t \bar t$ ZMF, 
and the factor 4 is conventional. The vectors
 ${\bf{\hat e}}_{i=1,2,3}$ in (\ref{eq:sbasis}) form an orthonormal basis.
The unnormalized expectation
values of these observables correspond to unnormalized double spin
asymmetries, i.e.,
to the following combination of $t, \bar t$ 
spin-dependent  cross sections: 
\begin{equation}
\label{double}
  \langle \langle {\cal O}_b\rangle
 \rangle_i
 \;  = \; \sigma_i (\uparrow \uparrow)+\sigma_i(\downarrow \downarrow)
  - \sigma_i(\uparrow \downarrow)- \sigma_i(\downarrow 
\uparrow) \, .
\label{dcorrw}
\end{equation}
The arrows on the right-hand side refer to the spin state of the top 
and antitop quarks  with respect to the  reference 
axes $ {\bf \hat a}$ and $ {\bf \hat b}$. 

\begin{figure}[H]
\begin{center}
\epsfig{file=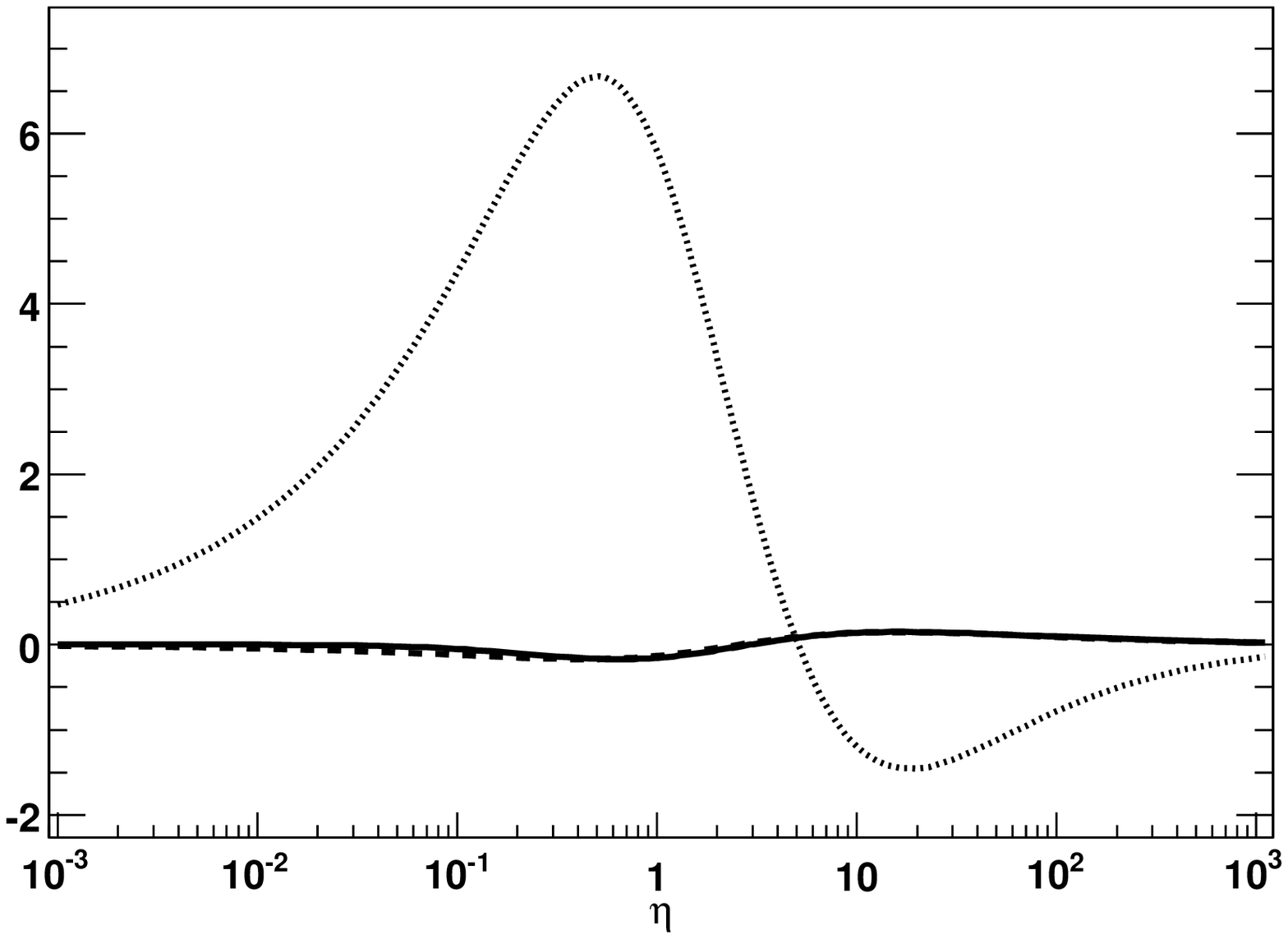, width=12cm,height=7.5cm }
\end{center}
\caption{Unnormalized helicity correlation
$\langle\langle {\cal O}_3 \rangle\rangle_{gg}$, defined in (\ref{eq:douevspin}), 
 in units of [pb]. The dotted line is the lowest order QCD contribution,
and the solid and dashed line is the weak contribution 
 for $m_H=120$ GeV and $m_H=200$ GeV, respectively.
} \label{fig:r3gg}
\end{figure}

\begin{figure}[H]
\begin{center}
\epsfig{file=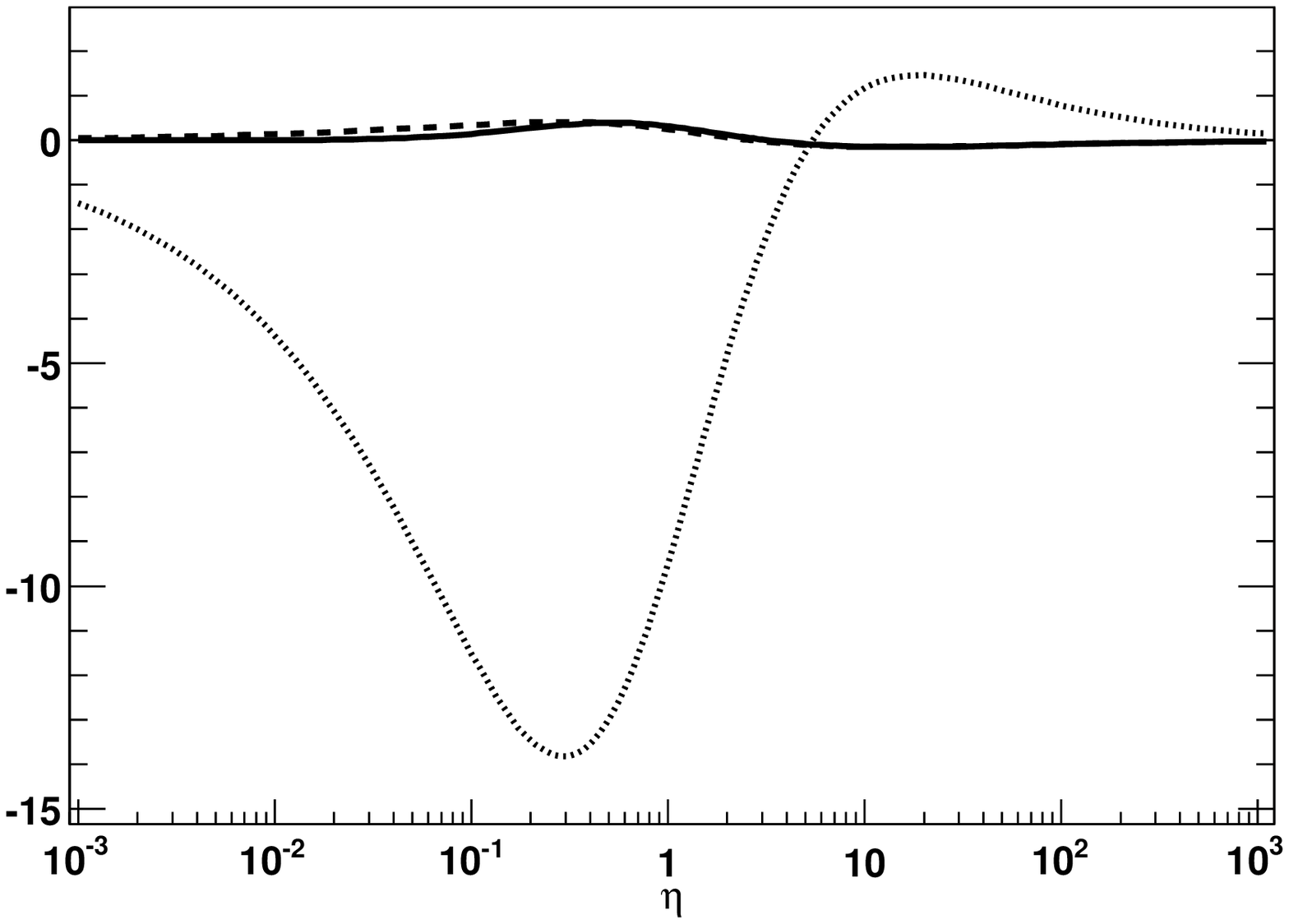, width=12cm,height=8cm }
\end{center}
\caption{Unnormalized spin correlation
$\langle\langle {\cal O}_4 \rangle\rangle_{gg}$, defined in (\ref{eq:douevspin}), 
 in units of [pb]. The dotted line is the lowest order QCD contribution,
and the solid and dashed line is the weak contribution 
 for $m_H=120$ GeV and $m_H=200$ GeV, respectively.
}\label{fig:r4gg}
\end{figure}

Again we compute the
order  $\alpha_s^2 \alpha$  weak contribution to (\ref{dcorrw})
and express it in terms of scaling functions. For
comparison we exhibit  also the lowest order  QCD term:
\begin{equation}
\langle \langle {\cal O}_b
\rangle {\rangle}_{gg} \, = \, 
\langle \langle {\cal O}_b
\rangle {\rangle}_{gg}^{(0)} +
\langle \langle {\cal O}_b
\rangle {\rangle}_{gg}^W  =\frac{4\pi}{m^2_t}[
\alpha_s^2 \, g^{(0,b)}_{gg}(\eta) + \alpha_s^2 \alpha \, g^{(1\, W,b)}_{gg}(\eta)]
\, .
\label{eq:douevspin}
\end{equation} 
The lowest order QCD and the weak contribution 
are plotted in Figs.~\ref{fig:r3gg} and \ref{fig:r4gg} for $b=3,4$,
respectively.
The NLO QCD contributions to (\ref{eq:douevspin}) were computed in 
\cite{Bernreuther:2004jv}. A comparison of the LO and NLO QCD and of the
weak contributions shows that, as far as the P-even
spin correlations are concerned, the SM weak interaction effects are
small 
in most of the  $t \bar t$ invariant mass range that is relevant for
the LHC. A closer inspection will be made in the next section.

For completeness we remark that  the absorptive parts of $\delta {\cal M}_W$ lead to T-odd
$t \bar t$ spin  correlations, both P-even and odd ones. These are 
very small effects, and we do not display them here. 
\section{Results for $pp$ ($p \bar p$) collisions}
Let us now investigate  $t \bar t$ production at the level of hadronic
collisions. Before
analyzing distributions, we first compute the weak corrections
to the hadronic $t \bar t$ cross section at the Tevatron and the LHC.
Table~1 contains the contributions
from $gg \to t {\bar t} (g)$, namely  at NLO QCD (that is, the 
sum of the first two terms in Eq.~(\ref{mixed})) and the weak
corrections of order $\alpha_s^2\alpha$, while Table~2 contains the
contributions from  $q {\bar q} \to t {\bar t} (g)$, using the results
for the weak corrections given in \cite{Bernreuther:2005is}.
 Notice that in this case the order $\alpha^2$ Born contribution must
 not be neglected as compared to the $\alpha_s^2\alpha$ term.
Here we use the  NLO parton distribution functions (PDF)
CTEQ6.1M \cite{Pumplin:2002vw}. The Tables show that the
weak interaction correction to the total cross section 
 is negative at the LHC and amounts to about $-1.3 \%$,
while it is about $0.5\%$ at the Tevatron.  These contributions 
are much smaller than the scale uncertainties of the fixed-order
NLO QCD corrections.

\begin{table}
\centering
\caption{
The  $gg$-induced hadronic
  $t \bar t$ cross section at the Tevatron ($\sqrt{s}=1.96$ TeV) and
at the LHC ($\sqrt{s}=14$ TeV) in
  units of pb, using the NLO parton distribution functions 
CTEQ6.1M \cite{Pumplin:2002vw},
$m_t$ = 172.7 GeV, two values of the Higgs mass,  and three different
values of $\mu$. We put $\mu\equiv \mu_R=\mu_F$.
\label{tab1gg}} 
\vspace*{0.5cm}
\begin{tabular}{|c|c|c|c|c|}\hline
\multicolumn{2}{|c|}{} & $\mu={m_t}/{2}$  & $\mu=m_t$  &
$\mu=2m_t$  \\ \hline
 Tevatron     & NLO QCD  &  1.293  & 1.107  & 0.891   \\ \cline{2-5}
               & weak  &  &   &      \\ 
              & $m_H=120$ GeV & $-0.0176$ & $-0.0111$  &  $-0.0073$ \\
              & $m_H=200$ GeV & $-0.0212$   & $-0.0135$  & $-0.0090$ \\ \hline
LHC  & & & &\\
             & NLO QCD & 794.544  & 769.988 & 712.341 \\ \cline{2-5}
              & weak    &   &  &  \\
              & $m_H=120$ GeV & $-13.137$  & $-10.095$ & $-7.892$ \\ 
              & $m_H=200$ GeV & $-13.511$  & $-10.431$ & $-8.198$ \\  \hline
\end{tabular}
\end{table}
\begin{table}
\centering
\caption{The  $q \bar q$-induced hadronic
  $t \bar t$ cross section at the Tevatron ($\sqrt{s}=1.96$ TeV) and
at the LHC ($\sqrt{s}=14$ TeV) in
  units of pb, using the NLO parton distribution functions 
CTEQ6.1M \cite{Pumplin:2002vw},
$m_t$ = 172.7 GeV,  $m_H$ = 120 GeV, and three different values of $\mu$.
 \label{tab1qq}} 
\vspace*{0.5cm}

\begin{tabular}{|c|c|c|c|c|}\hline
\multicolumn{2}{|c|}{ } & $\mu={m_t}/{2}$  & $\mu=m_t$  & $\mu=2m_t$ \\ \hline
 Tevatron     & NLO QCD  &  6.200 & 5.998  & 5.423  \\ \cline{2-5}
               & weak   &  0.0515 & 0.0466 & 0.0419 \\ \hline
LHC  & & & &\\
      & NLO QCD & 73.606 & 80.397 & 81.202 \\ \cline{2-5}
              &weak    & $-0.990$ & $-0.695$ & $-0.476$ \\
 \hline
\end{tabular}
\end{table}

In the remaining section we  study
 the weak interaction corrections for a  number of distributions and
compare them, in the case of  P-invariant observables, with 
the lowest order QCD
results. Therefore we compute  these distributions with 
the  LO parton distribution functions 
CTEQ6.L1 \cite{Pumplin:2002vw} which we take at the factorization scale $\mu=2
m_t$, and the input from the $gg$ and $q \bar q$ initiated
subprocesses  is evaluated
with the values of the QCD and QED couplings given in the previous
Section.

\begin{figure}[H]
\begin{center}
\epsfig{file=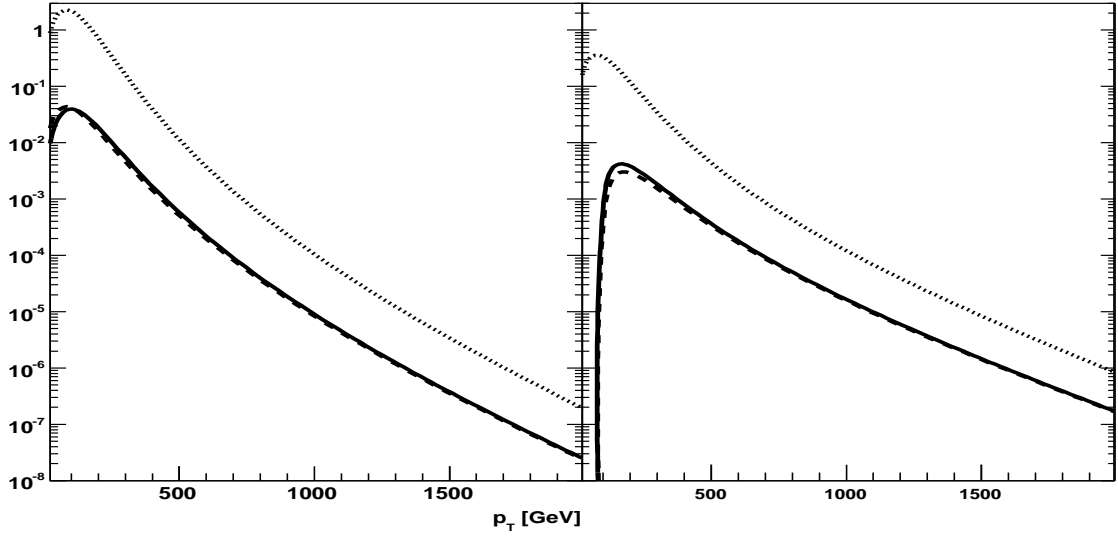, width=16cm,height=8cm }
\end{center}
\caption{Left frame: Contributions to the transverse top-momentum 
 distribution $d\sigma(gg)/dp_T$ at the LHC due to the $gg$
subprocess  in units of [pb/GeV].
 The dotted line is due to lowest order QCD,
and the solid and dashed line is the weak correction multiplied by -1
 for $m_H=120$ GeV and $m_H=200$ GeV, respectively. Right frame: The same 
for the $q \bar q$ process. 
}\label{fig:dsggdpt}
 \end{figure}

\begin{figure}[H]
\begin{center}
\epsfig{file=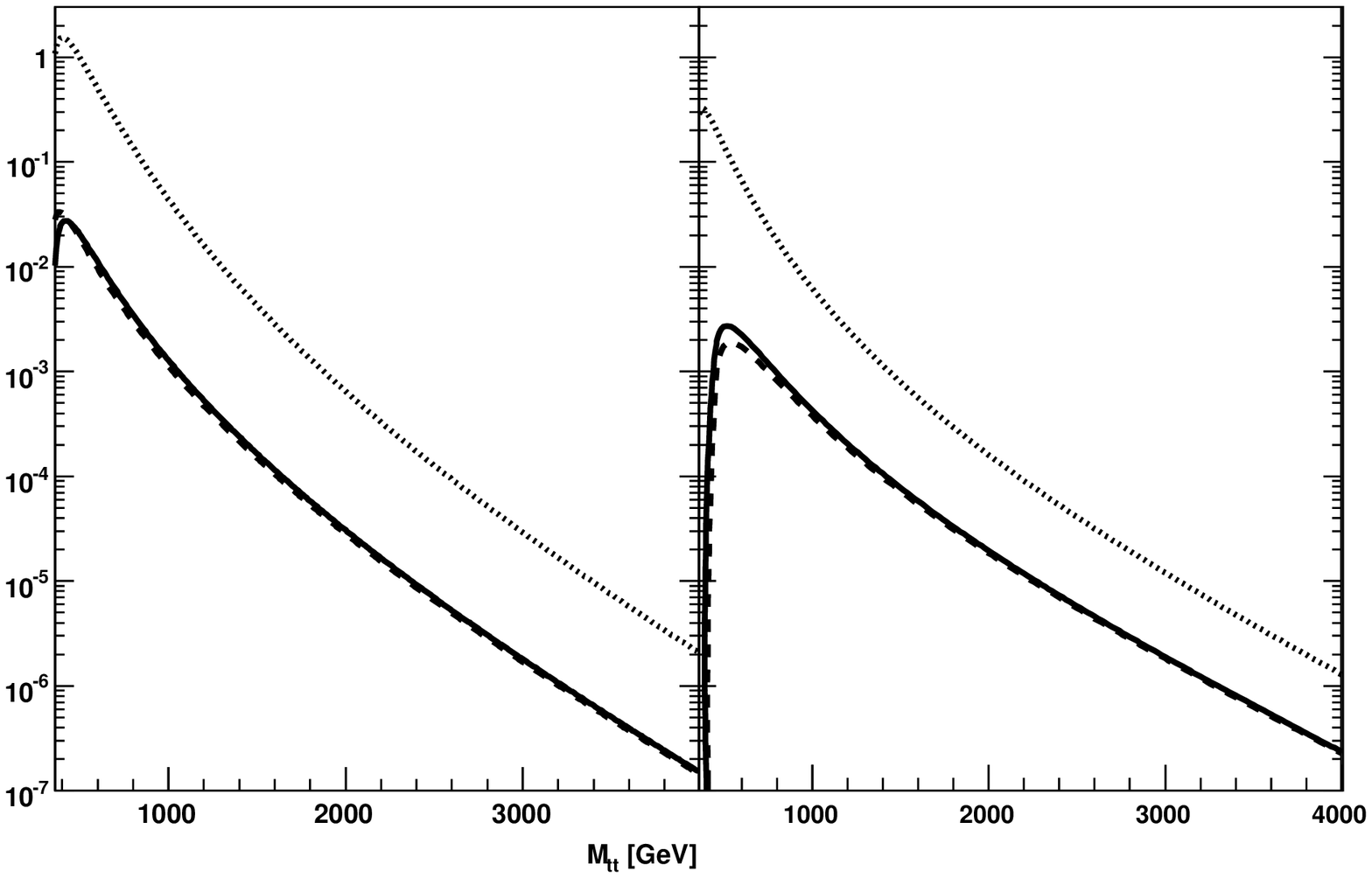, width=16cm,height=7cm }
\end{center}
\caption{Left frame: Contributions to the invariant mass
 distribution $d\sigma (gg)/ d\mtt$ at the LHC due to the $gg$
 subprocess in units of [pb/GeV]. The dotted line is due to lowest order QCD,
and the solid and dashed line is the weak correction multiplied by -1
 for $m_H=120$ GeV and $m_H=200$ GeV, respectively. Right frame: The
 same for the $q \bar q$ subprocesses.  
}\label{fig:dsggdm}
 \end{figure}

Figs.~\ref{fig:dsggdpt} and \ref{fig:dsggdm}, left frames, show 
the weak corrections for the $gg$ subprocess at the LHC,
multiplied by $-1$, to the tranverse top-momentum and to
the $t \bar t$ invariant mass  distribution, together with the lowest order QCD
results.  We have compared our results Figs.~\ref{fig:dsggdpt} and
\ref{fig:dsggdm} with those of \cite{KSU} and we 
agree. Comparing with the results of  \cite{Moretti:2006ea} shown
in Fig.~1 of that paper 
we disagree for $p_T\sim 100$ GeV and $\mtt\sim 400$ GeV, where
\cite{Moretti:2006ea} finds $d\sigma_{weak}$ to be positive below these
values, while we obtain that $d\sigma_{weak}$ is always negative for
the $gg$ subprocess and for $m_H \gtrsim 120$ GeV.

Figs.~\ref{fig:dsggdpt} and \ref{fig:dsggdm}, right frames, show 
the weak corrections for the $q \bar q$ subprocesses at the LHC,
multiplied by $-1$, to the tranverse top-momentum and to
the $t \bar t$ invariant mass  distribution, together with the lowest order QCD
results. The weak corrections consist of the ${\cal O}(\alpha^2)$ $q
{\bar q} \to \gamma, Z \to  t \bar t$ Born terms and the  ${\cal
  O}(\alpha_s^2 \alpha)$ virtual and real corrections to $q
{\bar q} \to  t {\bar t} (g)$. The latter corrections  are separately
infrared-divergent due to soft gluon radiation. We have computed these two
distributions with our results of
\cite{Bernreuther:2005is}, where  these divergences were treated
with a phase-space slicing procedure,  and we have checked that the sum of the virtual and real
corrections of ${\cal O}(\alpha_s^2 \alpha)$ are independent of the 
arbitrary slicing parameter $x_{cut}$ if it is small enough.
For $p_T \lesssim 100$ GeV, respectively $\mtt \lesssim 430$ GeV, the weak corrections are positive for the $q \bar q$
subprocesses.  The size of the ${\cal O}(\alpha^2)$ Born terms, which are of
course always positive,  are between $10\%$ and $20\%$  of the
${\cal O}(\alpha_s^2 \alpha)$ corrections in the $\mtt$
range considered, except in the vicinity of $\mtt \sim 450 $ GeV, where
the  ${\cal O}(\alpha_s^2 \alpha)$ corrections have a zero.
Notice that, choosing $m_H=120 \, (200)$ GeV,
the weak $q \bar q$ contributions are larger in magnitude than the ones from $gg$
for $p_T> 930 \,(690)$ GeV.

\begin{figure}[H]
\begin{center}
\epsfig{file=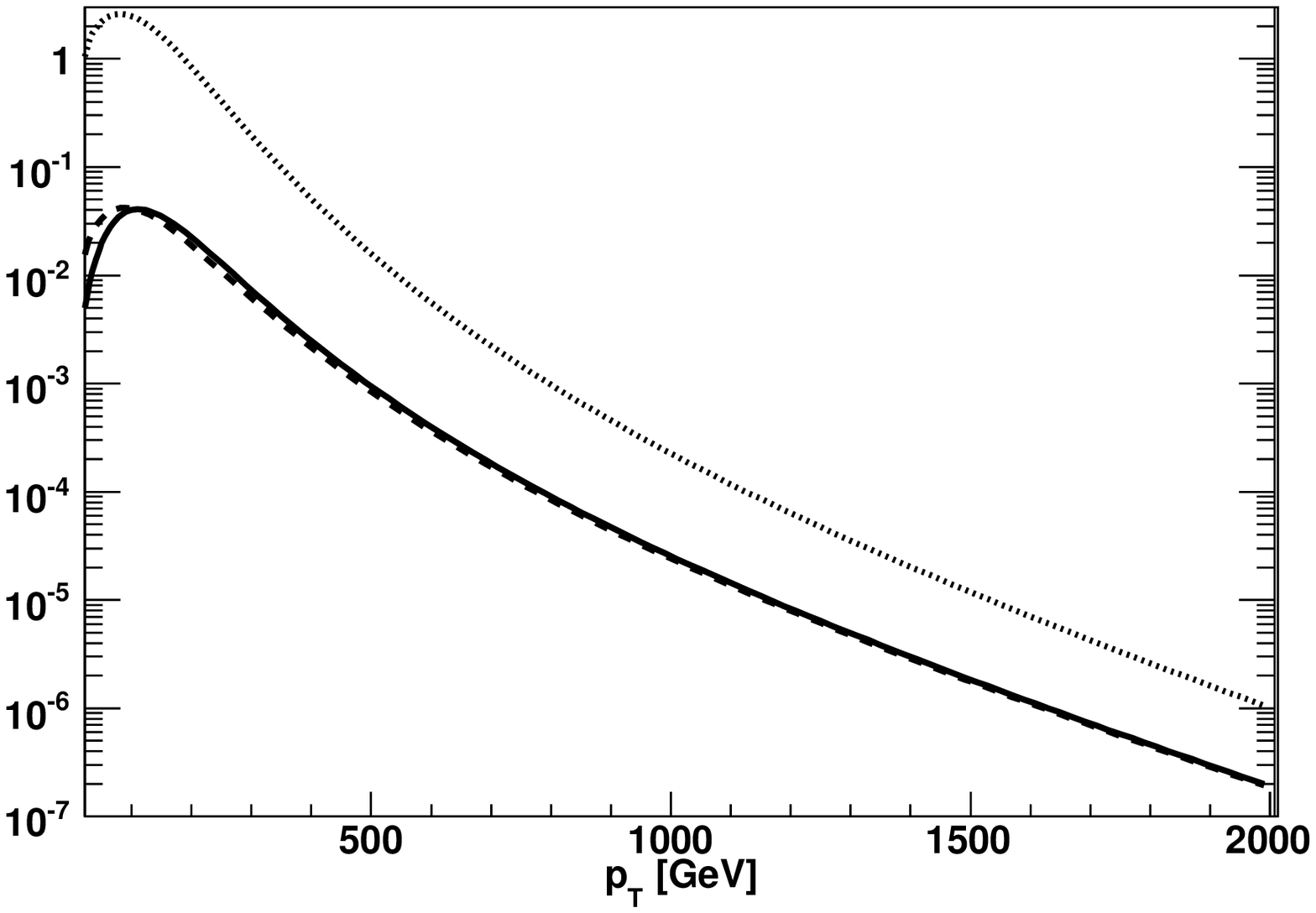, width=12cm,height=7.5cm }
\end{center}
\caption{Contributions to the transverse momentum
 distribution $d\sigma (gg+q{\bar q})/ dp_T$ at the LHC due to the $gg$
 and $q \bar q$ subprocesses in units of [pb/GeV].
 The dotted line is due to lowest order QCD,
and the solid and dashed line is the weak contribution multiplied by -1
 for $m_H=120$ GeV and $m_H=200$ GeV, respectively.
}\label{fig:sumdspt}
 \end{figure}

\begin{figure}[H]
\begin{center}
\epsfig{file=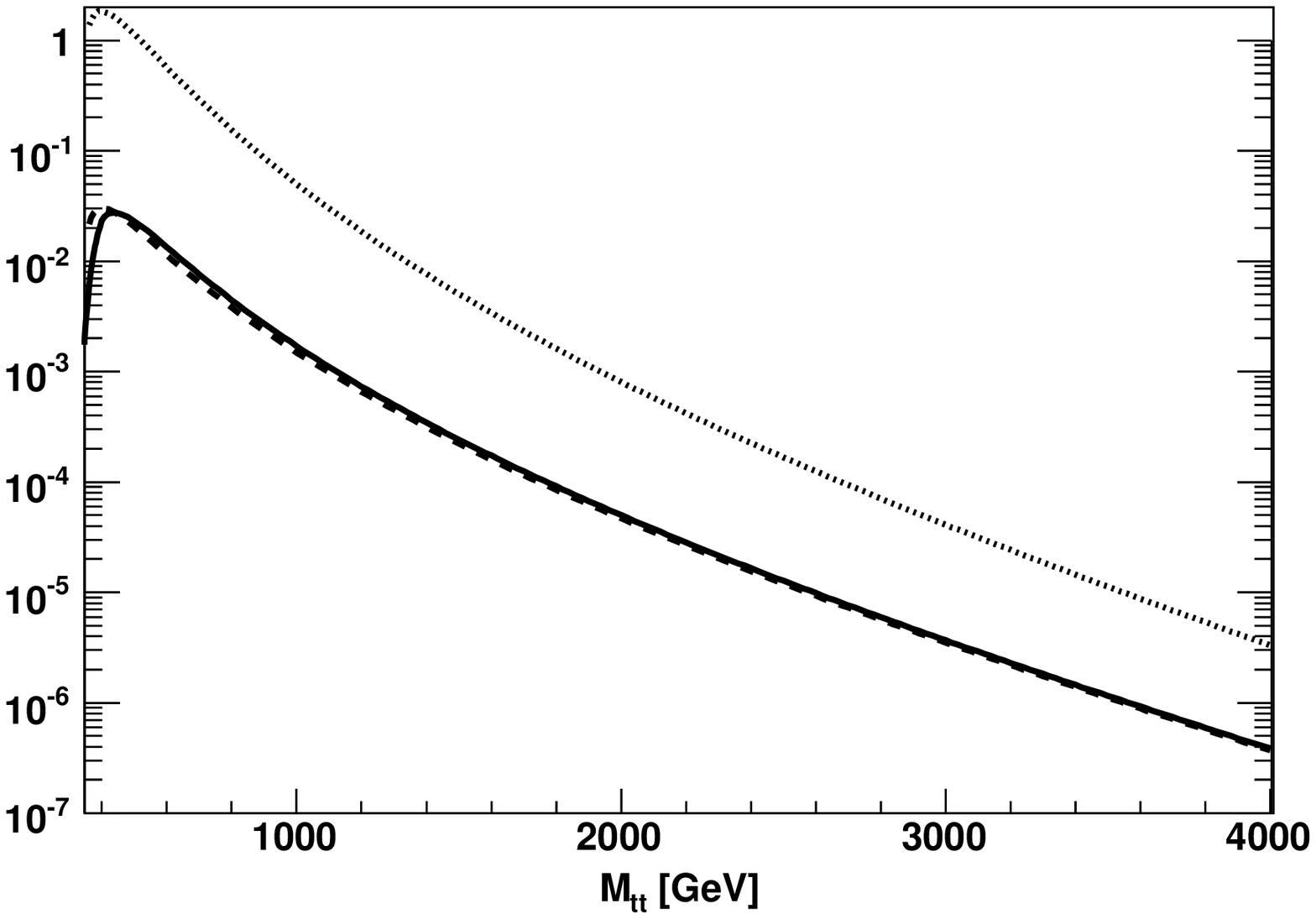, width=12cm,height=8cm }
\end{center}
\caption{Contributions to the invariant mass
 distribution $d\sigma (gg+q{\bar q})/ d\mtt$ at the LHC due to the $gg$
 and $q \bar q$ subprocesses in units of [pb/GeV].
 The dotted line is due to lowest order QCD,
and the solid and dashed line is the weak contribution multiplied by -1
 for $m_H=120$ GeV and $m_H=200$ GeV, respectively.
}\label{fig:sumdsdm}
 \end{figure}

\begin{figure}[H]
\begin{center}
\epsfig{file=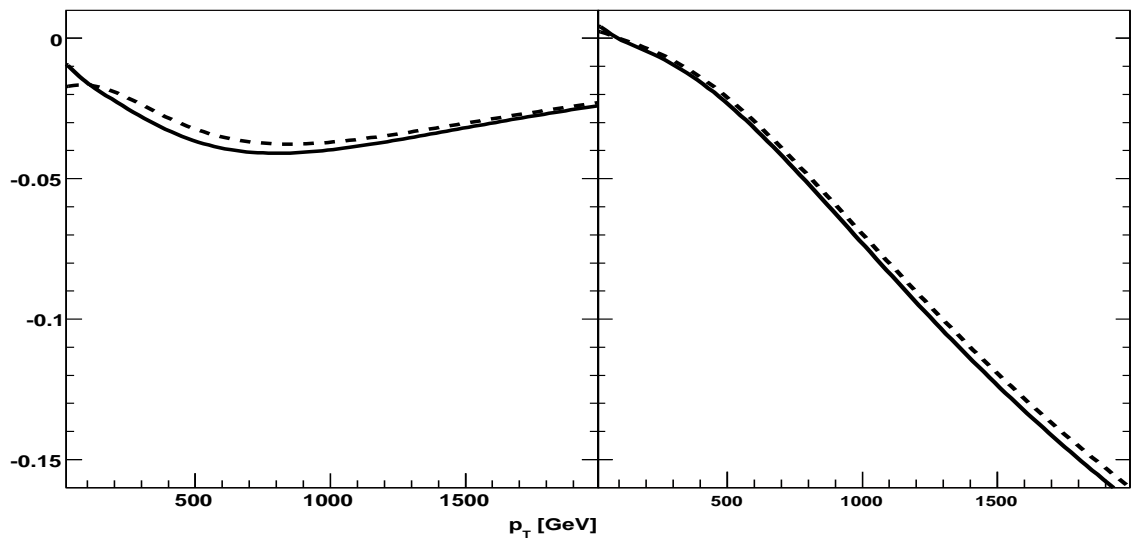, width=16cm,height=8cm }
\end{center}
\caption{Ratios of $d\sigma_{weak}(gg)/dp_T$ (left
  frame), $d\sigma_{weak}(q {\bar q})/dp_T$ (right 
  frame) and
  $d\sigma_{LO}(gg+q{\bar q})/ dp_T$ at the LHC.
The solid and dashed line is for $m_H=120$ GeV and $m_H=200$ GeV,
respectively. 
}\label{fig:ggqqpthr}
 \end{figure}

\begin{figure}[H]
\begin{center}
\epsfig{file=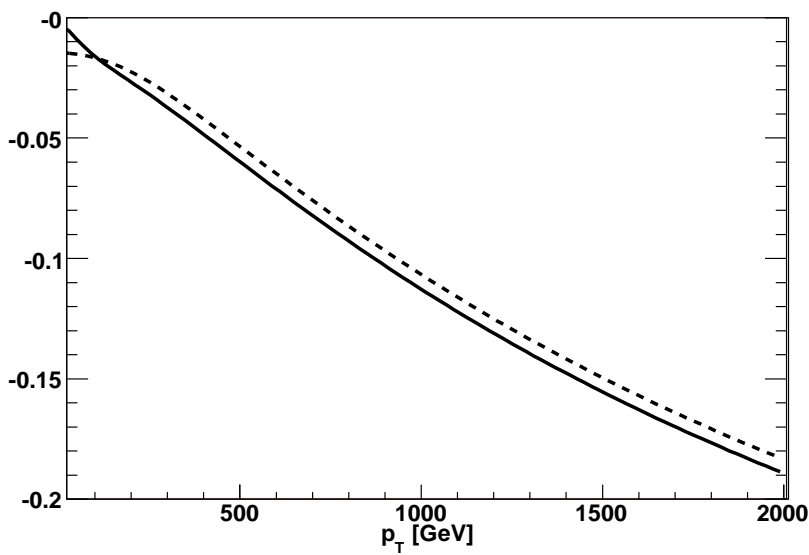, width=12cm,height=8cm }
\end{center}
\caption{Ratio of $d\sigma_{weak}(gg+q{\bar q})/dp_T$ and
  $d\sigma_{LO}(gg+q{\bar q})/dp_T$ at the LHC.
The solid and dashed line is for $m_H=120$ GeV and $m_H=200$ GeV, respectively.
}\label{fig:sumpthr}
 \end{figure}

\begin{figure}[H]
\begin{center}
\epsfig{file=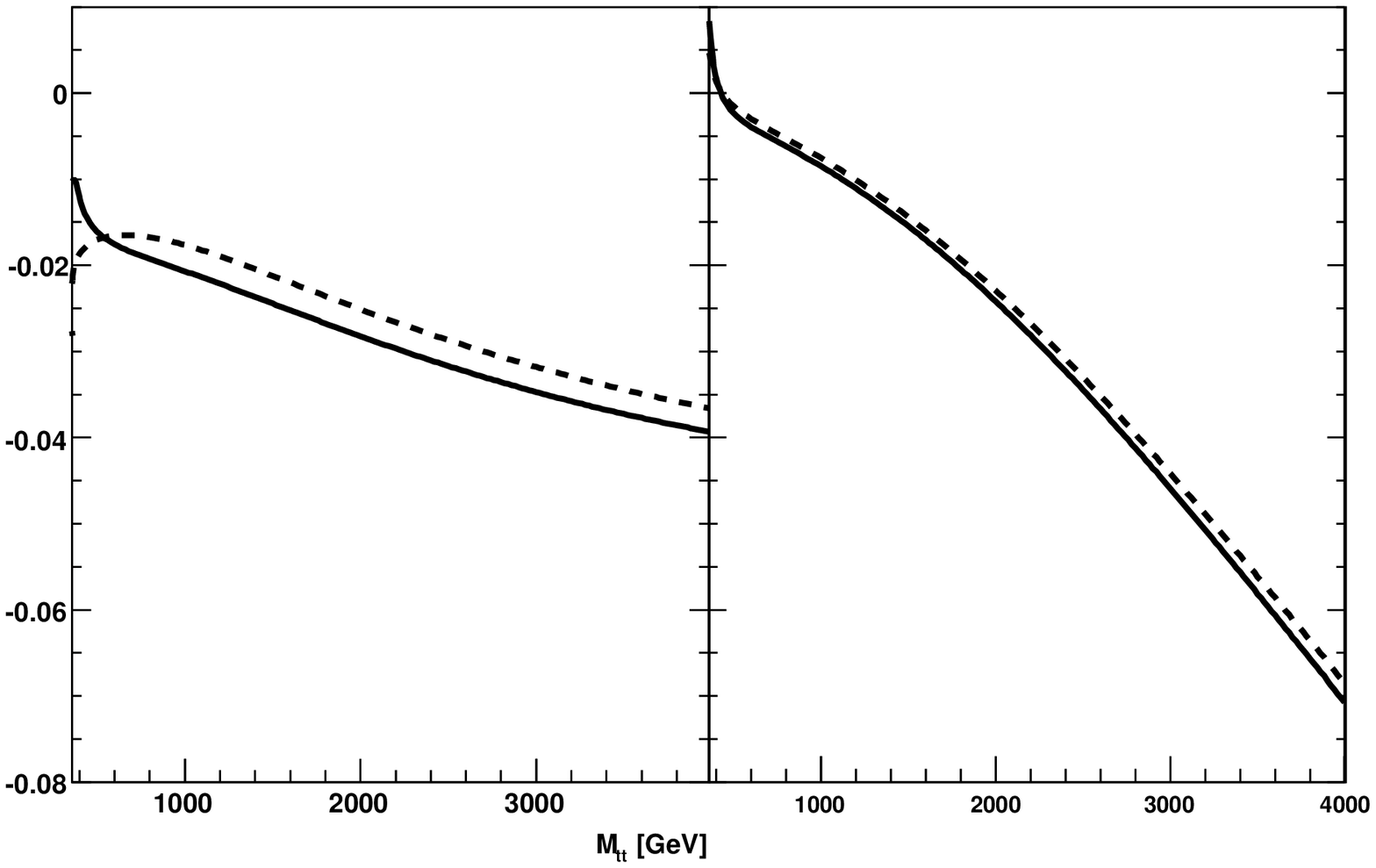, width=16cm,height=8cm }
\end{center}
\caption{Ratios of $d\sigma_{weak}(gg)/ d\mtt$ (left
  frame), $d\sigma_{weak}(q {\bar q})/ d\mtt$ (right 
  frame) and
  $d\sigma_{LO}(gg+q{\bar q})/ d\mtt$ at the LHC.
The solid and dashed line is for $m_H=120$ GeV and $m_H=200$ GeV,
respectively. 
}\label{fig:summttr}
 \end{figure}

\begin{figure}[H]
\begin{center}
\epsfig{file=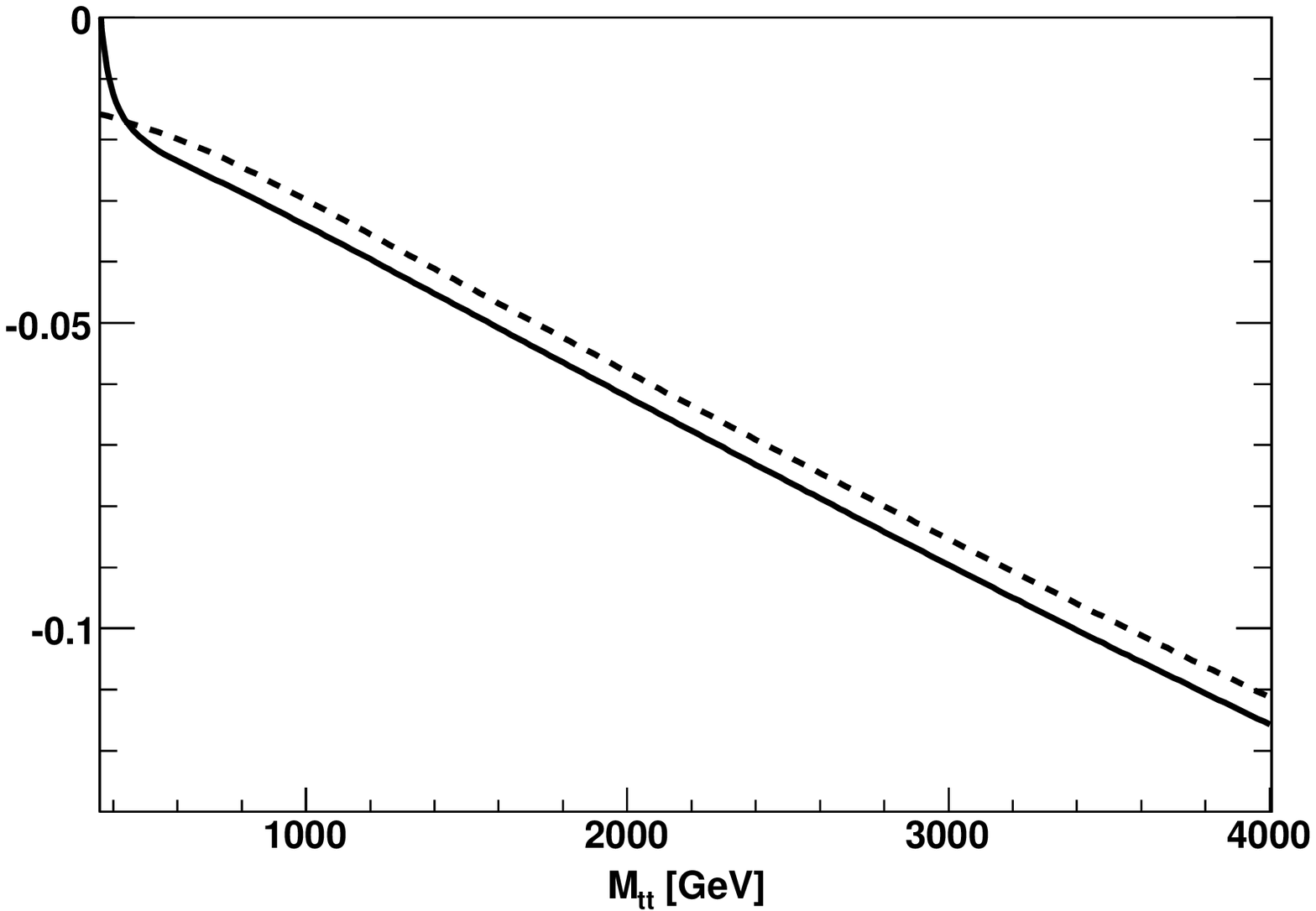, width=12cm,height=8cm }
\end{center}
\caption{Ratio of $d\sigma_{weak}(gg+q{\bar q})/ d\mtt$ and
  $d\sigma_{LO}(gg+q{\bar q})/ d\mtt$ at the LHC.
The solid and dashed line is for $m_H=120$ GeV and $m_H=200$ GeV, respectively.
}\label{fig:sumcshr}
 \end{figure}

Figs.~\ref{fig:sumdspt} and~\ref{fig:sumdsdm}  show the sum of the weak corrections to the
transverse top-momentum and the
$t \bar t$ invariant mass distribution, together with the LO QCD results.
Figs.~\ref{fig:ggqqpthr} and~\ref{fig:sumpthr} show 
the ratios \\
$[d\sigma_{weak}(gg)/ dp_T]/[d\sigma_{LO}/ dp_T]$, 
$[d\sigma_{weak}(q{\bar q})/ dp_T]/[d\sigma_{LO}/ dp_T]$, 
and the sum, $[d\sigma_{weak}/ dp_T]/[d\sigma_{LO}/dp_T]$ for
the LHC, for two values of the Higgs mass, 
where $d\sigma_I= d\sigma_I(gg+q{\bar q})$, $I= weak, \, LO$. The
plots display clearly that 
the $q \bar q$ part of the weak corrections is not
negligible compared to those for the $gg$ subprocess; as already
mentioned, 
the $q \bar q$ contributions dominate for large $p_T$. 
In Figs.~\ref{fig:summttr} and~\ref{fig:sumcshr} the analogous ratios are displayed for the
$\mtt$ distribution.
Notice that in these ratios  the changes of $d\sigma_{weak}$
and $d\sigma_{LO}$ due to 
variations of the LO PDF and the LO QCD coupling with $\mu$ cancel to
a large extent.

\begin{figure}[H]
\begin{center}
\epsfig{file=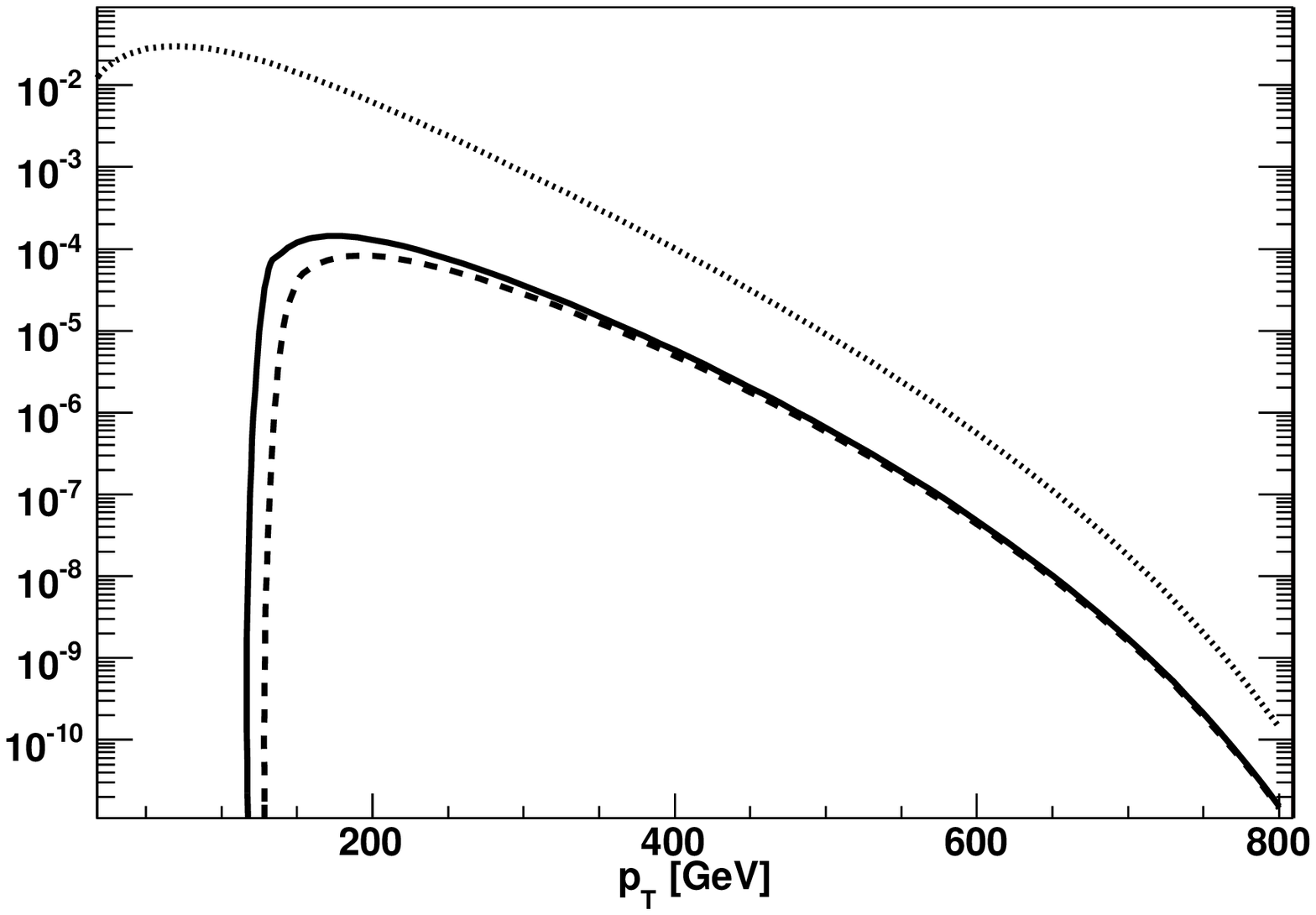, width=12cm,height=8cm }
\end{center}
\caption{Contributions to the transverse momentum
 distribution $d\sigma (gg+q{\bar q})/ dp_T$ at the Tevatron due to the $gg$
 and $q \bar q$ subprocesses in units of [pb/GeV].
 The dotted line is due to lowest order QCD,
and the solid and dashed line is the weak contribution multiplied by -1
 for $m_H=120$ GeV and $m_H=200$ GeV, respectively.
}\label{fig:tevsumdsdpt}
 \end{figure}

\begin{figure}[H]
\begin{center}
\epsfig{file=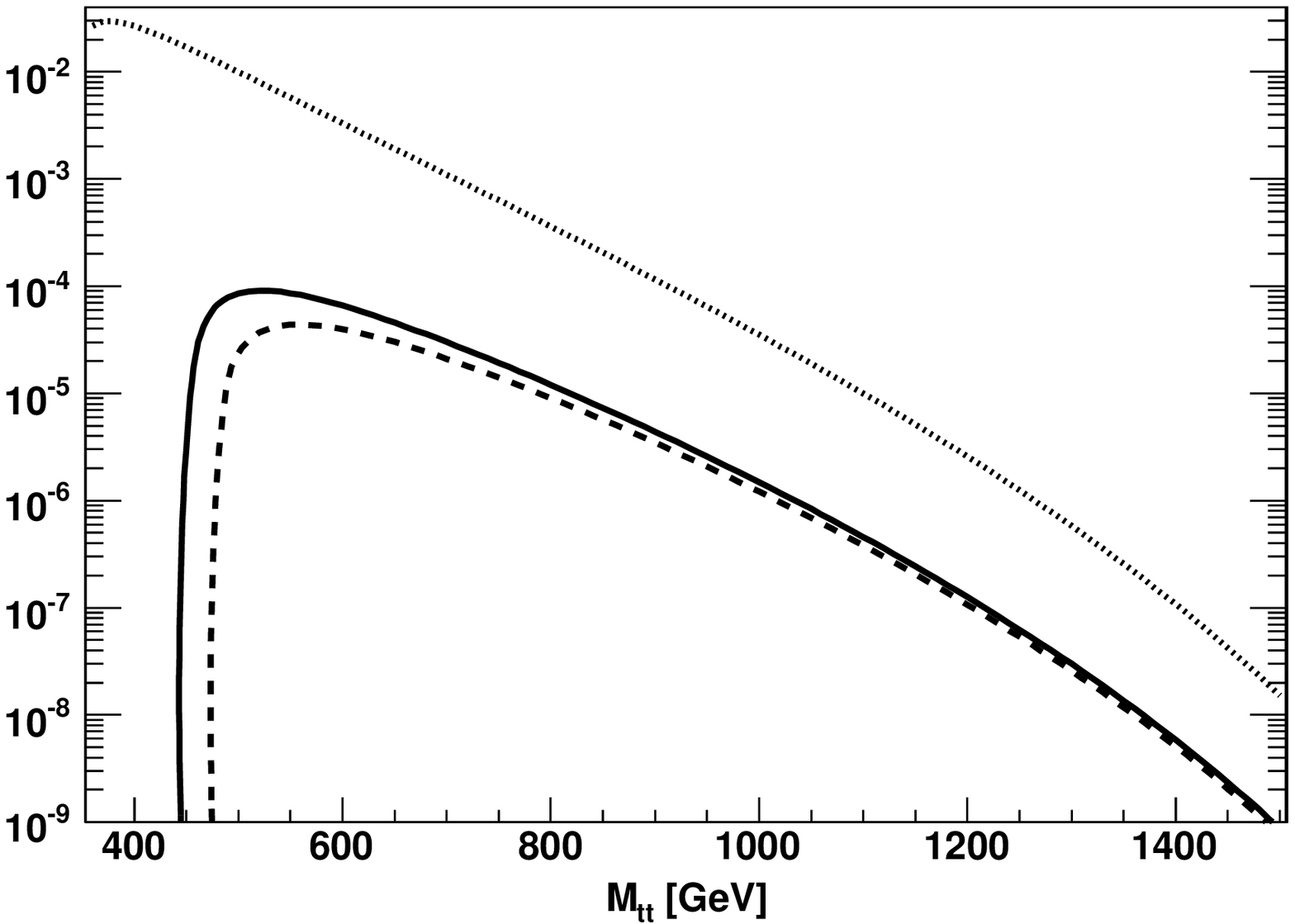, width=12cm,height=7cm }
\end{center}
\caption{Contributions to the invariant mass
 distribution $d\sigma (gg+q{\bar q})/ d\mtt$ at the Tevatron due to the 
 $q \bar q$ and $gg$ subprocesses in units of [pb/GeV].
 The dotted line is due to lowest order QCD,
and the solid and dashed line is the weak contribution multiplied by -1
 for $m_H=120$ GeV and $m_H=200$ GeV, respectively.
}\label{fig:tevsumdsdm}
 \end{figure}

\begin{figure}[H]
\begin{center}
\epsfig{file=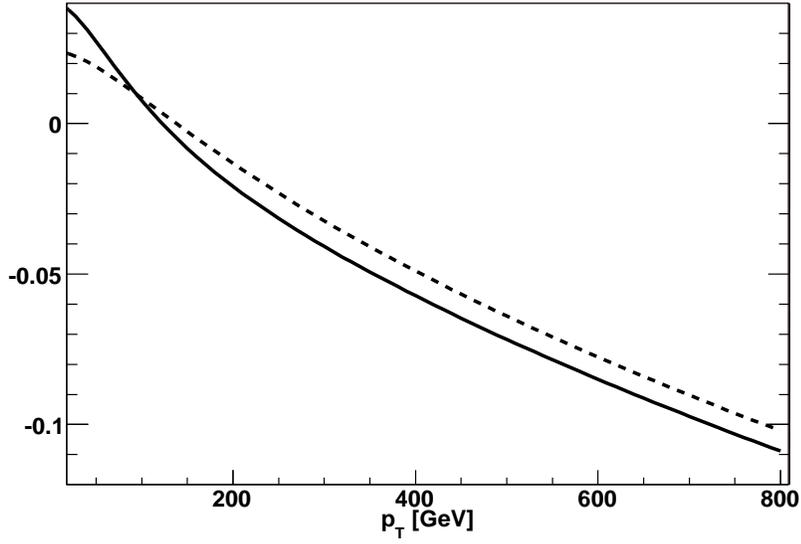, width=12cm,height=8cm }
\end{center}
\caption{Ratio of the distributions $(d\sigma/dp_T)_{weak}$
and  $(d\sigma/dp_T)_{LO, \, QCD}$, shown 
in Fig.~\ref{fig:tevsumdsdpt}, at the Tevatron
for  $m_H=120$ GeV (solid) and $m_H=200$ GeV (dashed).
}\label{fig:ptratioTeV}
\end{figure}

\begin{figure}[H]
\begin{center}
\epsfig{file=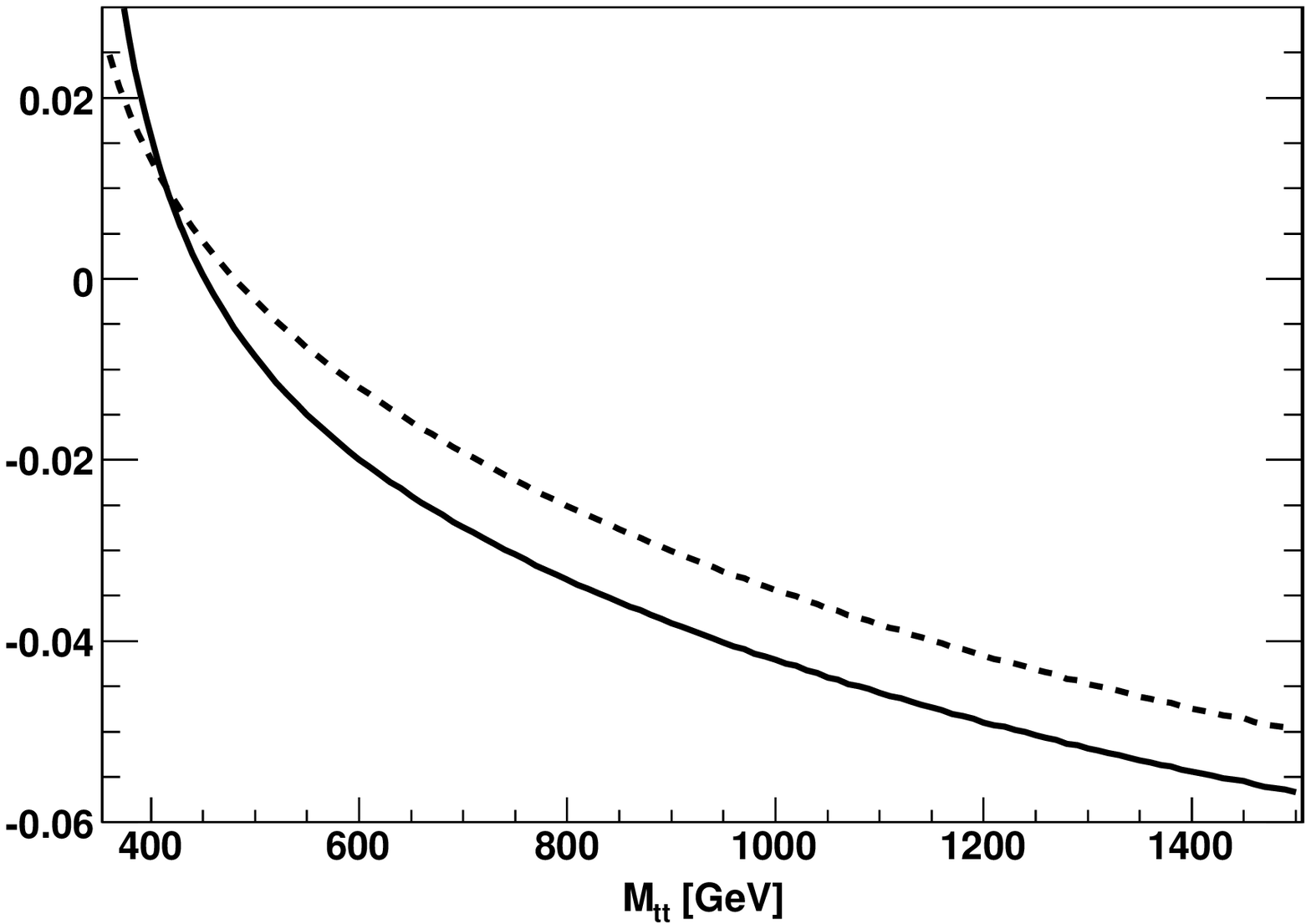, width=12cm,height=8cm }
\end{center}
\caption{Ratio of $d\sigma_{weak}(gg+q{\bar q})/ d\mtt$ and
  $d\sigma_{LO}(gg+q{\bar q})/ d\mtt$, shown 
in Fig.~\ref{fig:tevsumdsdm},  at the Tevatron.
The solid and dashed line is for $m_H=120$ GeV and $m_H=200$ GeV, respectively.
}\label{fig:mtratioTeV}
 \end{figure}

Figs.~\ref{fig:tevsumdsdpt}  and~\ref{fig:tevsumdsdm} display the weak and LO
QCD contributions to the $p_T$ and $\mtt$ distribution for the Tevatron, and 
Figs.~\ref{fig:ptratioTeV}  and~\ref{fig:mtratioTeV} show the corresponding ratios.                  
Here the weak corrections become positive for $p_T \lesssim 120$ GeV, 
$\mtt \lesssim$ 450 GeV if $m_H=120$ GeV.

The weak corrections to the distributions grow for large
$p_T$ and $\mtt$ as compared to the lowest order QCD results.
For the $p_T$ distribution at the LHC the sum of the weak corrections amounts
to about $-10\%$ for $p_T= 890\, (950)$ GeV for $m_H=120 \, (200)$ GeV. 
The weak corrections are 
less pronounced in the $\mtt$ distribution: for $\mtt = 2$ TeV 
and $m_H=120$ GeV they are about $-6 \%$
 as compared to the lowest order result.
Table~\ref{tabstatsig} contains, for the LHC and the Tevatron,
 the LO $t \bar t$ cross section and the weak
corrections for $\mtt$ and $p_T$ larger than a selected set of values
$p_T^*$ and $\mtt^*$, respectively. Furthermore the ratio $r=weak/LO$ and
the statistical significance $S=|r |{\sqrt N_{event}}$ in standard
deviations (s.d.) are tabulated  assuming an integrated
luminosity of 10 fb$^{-1}$ for the Tevatron and 100 fb$^{-1}$ for the LHC.

\begin{table}[ht]
\caption{Upper part: Cross section for $t \bar t$
events with $t \bar t$ invariant mass larger than $\mtt^*$ and for
events with $p_T > p_T^*$. The rows
contain the leading
order QCD  cross section and the weak corrections for two different
Higgs boson masses in units of pb, the respective ratios $r=weak/LO$ and the
statistical significance $S$ in s.d. assuming an
integrated
luminosity of 10 fb$^{-1}$ for the Tevatron and 100 fb$^{-1}$ for the
LHC.
The numbers for the LO cross section and the weak corrections given in
the table were rounded, while $r$ and $S$ were computed with the
precise numbers.
} \label{tabstatsig} 
\begin{center}
\begin{tabular}{|c|c|c|c|c|c|c|c|c|}
\hline
 &
 \multicolumn{8}{c|}{$\sigma(M_{t\overline{t}}>M_{t\overline{t}}^*)$
   \,  [pb] } \\
\hhline{~--------}
 & $\mtt^* \, \, \left[{\rm GeV}\right]$ & LO & weak, $m_H=120$ GeV  & r \, \, $\left[\%\right]$ &S  & weak,
 $m_H=200$ GeV & r \, \,  $\left[\%\right]$  & S\\
\hline
LHC & 500 &174.7  &$-4.42$& $-2.5$& 105& $-3.8$&$-2.2$ & 90 \\
\hhline{~--------}
    & 1000 &10.7 & $-0.44$& $-4.1$& 42 & $-0.39$&$-3.7$& 38\\
\hhline{~--------}
    & 1500 & 1.39& $-0.078$&$-5.6$ & 21 & $-0.073$&$-5.2$ &19\\
\hline
Tev. & 400 &2.65 &$-0.01$&$-0.4$ &0.6& 0.0009&0.03&0.05\\
\hhline{~--------}
    & 700 &0.098 & $-0.0031$& $-3.2$&1.0 & $-0.0023$&$-2.4$&0.75\\
\hhline{~--------}
    & 1000 &0.003 &$-0.00012$&$-4.5$ &0.23 & $-0.0001$&$-3.7$&0.2\\
\hhline{=========}
 &  \multicolumn{8}{c|}{$ \sigma(p_T>p_T^*)$ [pb] } \\
\hhline{~--------}
& $p_T^* \, \, \left[{\rm GeV}\right]$ &  LO &weak, $m_H=120$ GeV & r \, \,
$\left[\%\right]$  &S &  
weak,
$m_H=200$ GeV & r \, \,  $\left[\%\right]$  & S\\
\hline
LHC&200 & 59.8   & $-2.1$  & $-3.5$ &85 &$-1.77$   & $-3.0$ & 72\\ \hhline{~--------}
   &500 & 1.6    & $-0.12$ & $-7.3 $ &29 & $-0.11$ & $-6.6$ & 27\\ \hhline{~--------}
   &1000 & 0.037 & $-0.0047$ & $-12.8$  &7 & $-0.004$ & $-12.2$ & 7\\ 
\hline
Tev.&100 &5.8 & $-0.016$   & $-0.28$ & 0.6   & $-0.007$  & $-0.1$ & 0.3\\\hhline{~--------}

   &200 &0.71 & $-0.01$   & $-1.4$ & 1.2   &  $-0.007$ & $-1.0$ & 0.8\\\hhline{~--------}

   &500 &0.011 &$ -0.00002$ & $-0.23$ &0.02  & $-0.00002$ & $-0.2$ & 0.02\\ 

\hline
\hline
\end{tabular}
\end{center}
\end{table}

The numbers for $S$ should be taken only as order of magnitude
estimates because, as emphasized in the previous section, 
a precise determination $r$ of these ratios must
take the NLO QCD corrections into account, both near threshold and
for large $p_T$ or $\mtt$, where these corrections are dominant. 
A definite statement about whether or not SM weak interaction effects
are ``visible'' in distributions like $d\sigma/dp_T$ and
$d\sigma/d\mtt$ for large $p_T$, $\mtt$ requires a reliable
computation of $d\sigma_{QCD}$ beyond the leading
order, including resummation of gluon radiation
\cite{Bonciani:1998vc,Kidonakis:2001nj,Cacciari:2003fi,Kidonakis:2004hr,Banfi:2004xa}
and a study of the renormalization and factorization scale uncertainties in that
range. This is an important issue, especially
 at the LHC, as the $\mtt$ spectrum probes the
existence of exotic heavy resonances that strongly couple to $t \bar t$,
but is beyond the scope of this paper.

\begin{figure}[H]
\begin{center}
\epsfig{file=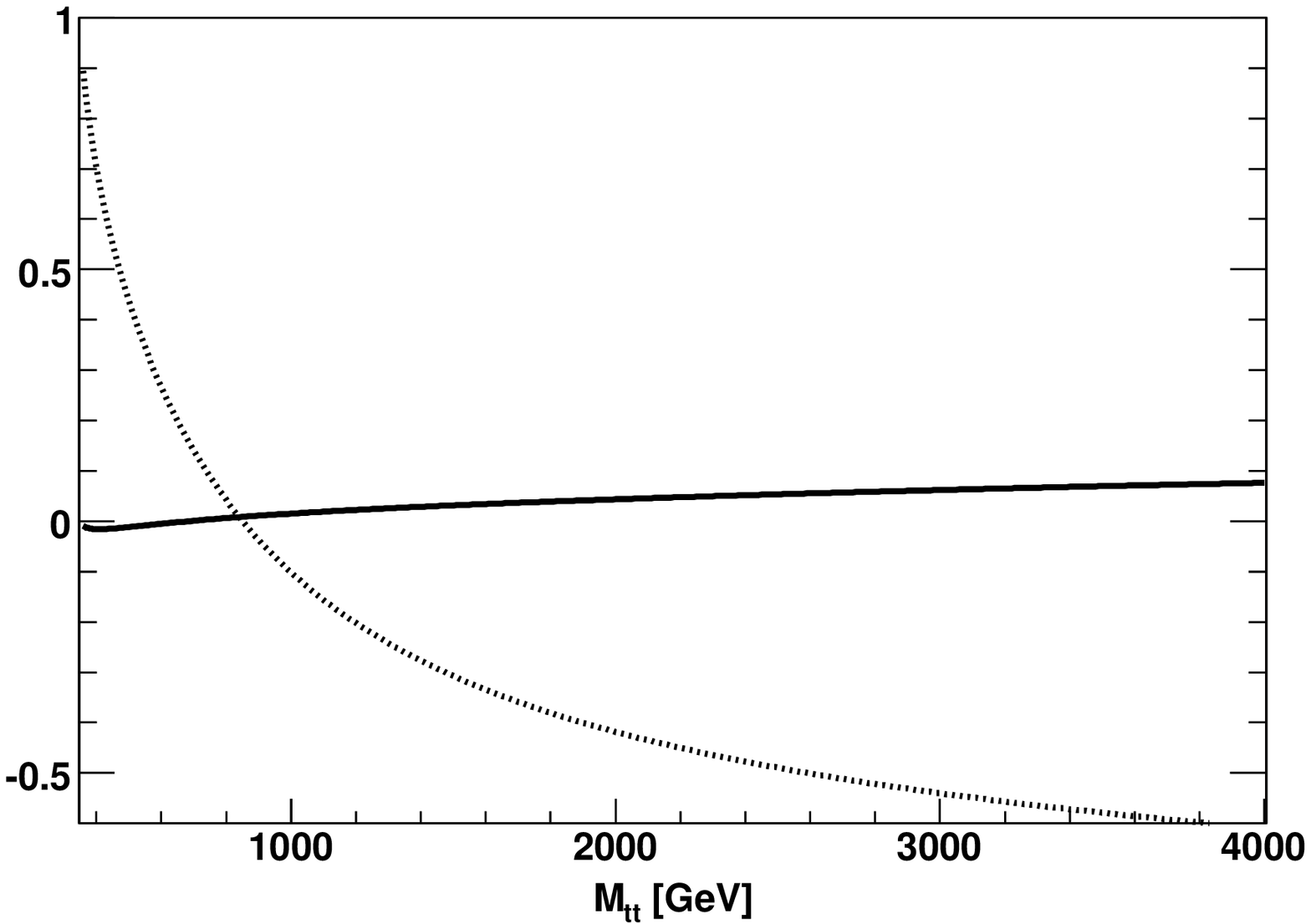, width=12cm,height=7cm }
\end{center}
\caption{The P-invariant differential double spin 
asymmetry $A_{hel}$, defined in (\ref{ashel}), at the LHC ($gg$
subprocess only). 
 The dotted and solid line is the contribution from lowest order QCD
and from weak interactions with $m_H=120$ GeV, respectively. Using  $m_H=200$ GeV
does not lead to a significant change of the solid line.
}\label{fig:Adshel}
 \end{figure}

\begin{figure}[H]
\begin{center}
\epsfig{file=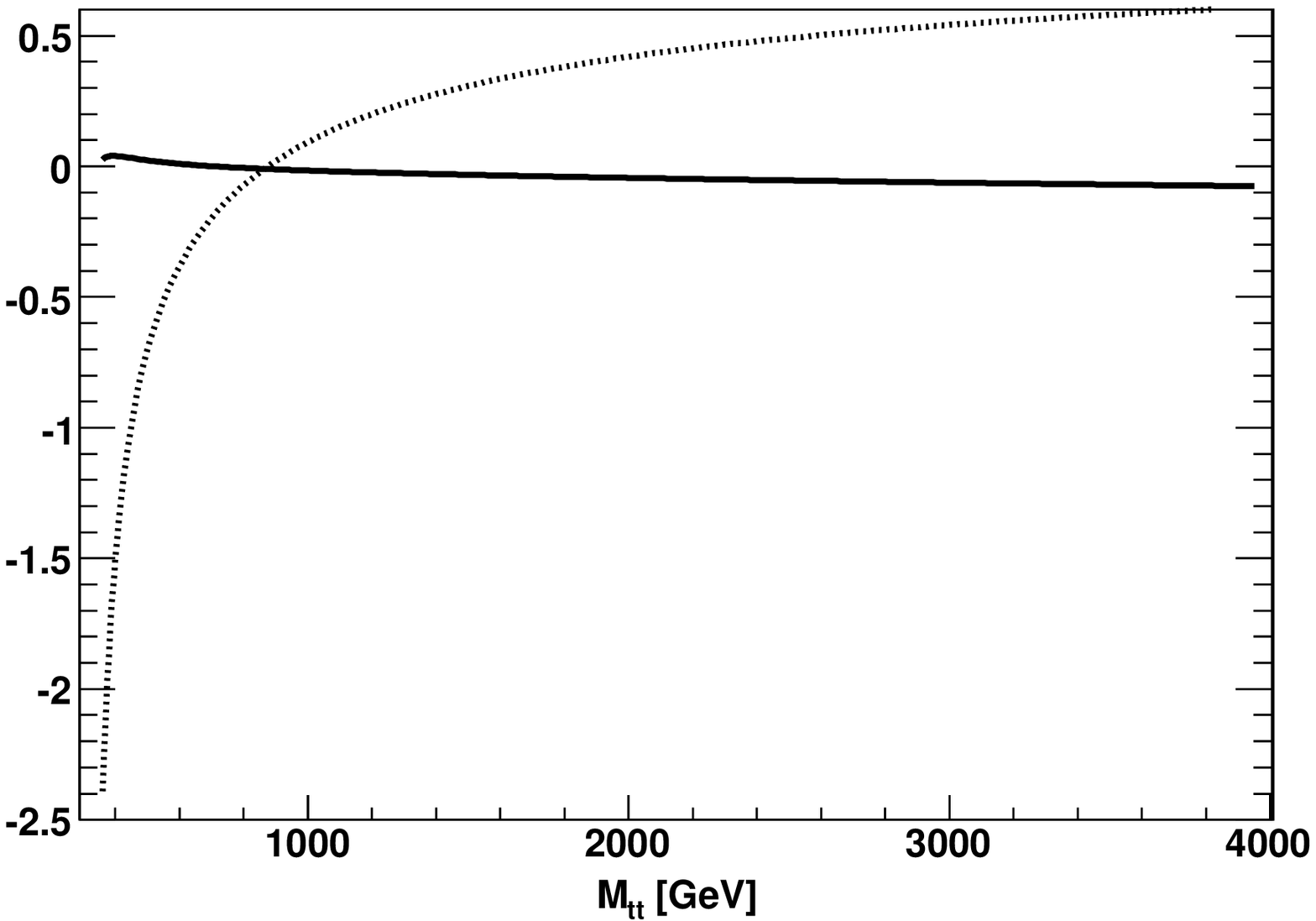, width=12cm,height=7cm }
\end{center}
\caption{The P-invariant differential double spin 
asymmetry $A_{spin}$  at the LHC ($gg$
subprocess only). The dotted and solid line is the contribution from lowest order QCD
and from weak interactions with $m_H=120$ GeV, respectively.
 Using  $m_H=200$ GeV
does not lead to a significant change of the solid line.
}\label{fig:Adspin}
 \end{figure}

\begin{figure}[H]
\begin{center}
\epsfig{file=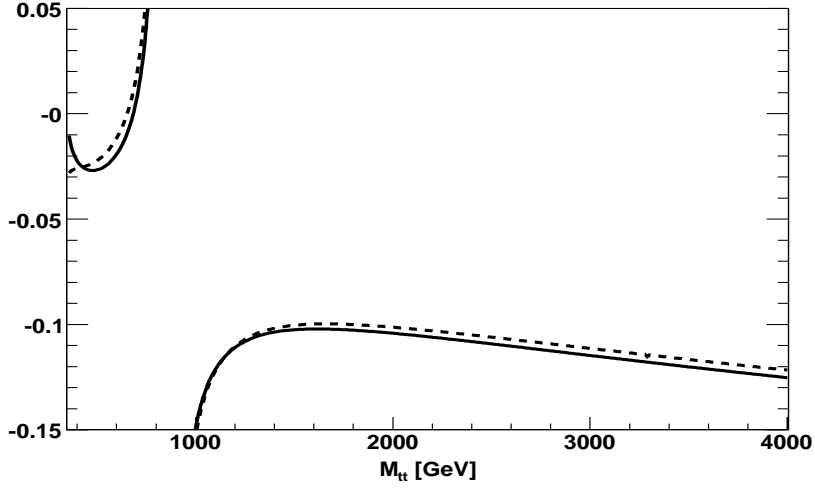, width=12cm,height=7cm }
\end{center}
\caption{The ratio  $A_{hel}^{weak}/A_{hel}^{LO}$, 
of the contributions to the double spin asymmetry
$A_{hel}$ for  the LHC ($gg$
subprocess only). 
 The solid and dotted line corresponds to  $m_H=120$ GeV and
$m_H=200$ GeV, respectively.
}\label{fig:Adshelrat}
 \end{figure}

Next we consider two differential  parity-conserving  double spin asymmetries
for the LHC which correspond to the spin-spin correlation
observables ${\cal O}_3$ and  ${\cal O}_4$ above. For brevity only the
contribution from the $gg$ subprocess will be taken into account, see
\cite{Bernreuther:2005is} for the $q \bar q$ contributions. 
The helicity correlation ${\cal O}_3$ leads to the
asymmetry $d\sigma_{++}+d\sigma_{--}-d\sigma_{+-}- d\sigma_{-+}$,
where as before the first (second) subscript refers to the $t$ $({\bar t})$
helicity. With $N_g\equiv d\sigma(gg)/d\mtt$, where $d\sigma(gg)$
denotes the LO QCD and the weak contributions for the $gg$ subprocess,
we consider
\begin{equation}
A_{hel} \equiv N_g^{-1}\left( {\frac{d\sigma_{++}}{d\mtt}+ \frac{d\sigma_{--}}{d\mtt} 
- \frac{d\sigma_{+-}}{d\mtt}- \frac{d\sigma_{-+}}{d\mtt}} \right)  \, ,
\label{ashel}
\end{equation}
In complete analogy to $A_{hel}$ we can define an asymmetry
$A_{spin}$ based on the spin-spin projection ${\cal O}_4$. For
numerical evaluation we decompose the numerator of  (\ref{ashel}) into
the  QCD Born and the weak contribution, 
$A_{hel} = A_{hel}^{LO}+ A_{hel}^{weak}$. The same decomposition is made for $A_{spin}$. 
The two pieces are shown as functions
of $\mtt$ for the helicity and spin asymmetry in
Figs.~\ref{fig:Adshel} and \ref{fig:Adspin}, respectively.
In Fig.~\ref{fig:Adshelrat} the ratio $A_{hel}^{weak}/A_{hel}^{LO}$
is plotted. The corresponding ratio for $A_{spin}$, which is not displayed
here, looks almost identical. Fig.~\ref{fig:Adshelrat} shows that the
weak corrections are about $-10 \%$ of the Born term 
for $\mtt > 1$ TeV. Near $\mtt = 900$ GeV, $A^{LO}$
has a zero. However, the 
NLO QCD corrections to these correlations, computed in  \cite{Bernreuther:2004jv},
render $A^{QCD}$  non-zero at these values
of $\mtt$. For large $\mtt$ the NLO QCD corrections 
are the dominant contributions. Thus we conclude that
the SM weak interaction contributions to parity-invariant 
double spin asymmetries at the LHC are quite small. Nevertheless,
they should be taken into account in SM predictions  in view of the estimated 
error of about 5$\%$
with which these asymmetries may be measured \cite{Hubaut:2005er}.

The quantity $A_{hel}$  and the ratio $A_{hel}^{weak}/A_{hel}^{LO}$ was also 
computed in \cite{Moretti:2006nf}. However, we disgree with the results given in
Figs.~2 and 3 of that paper. While we obtain that  $A_{hel}^{weak}/A_{hel}^{LO}
\lesssim -0.1$ for $\mtt \gtrsim 1.4$ TeV (c.f. Fig.~\ref{fig:Adshelrat}),
the corresponding result in \cite{Moretti:2006nf} is much smaller in magnitude.

Finally we analyze the P-violating single and double top-spin
asymmetries of Section 2 at the level of hadronic collisions. 
The double spin asymmetries (\ref{dupdown}) for the various parton initial states $i$
lead to the differential asymmetry
\begin{equation}
\Delta(\uparrow\downarrow) \,\equiv \,
N^{-1}\left( \frac{d\sigma(\uparrow\downarrow )}{d\mtt} - 
\frac{d\sigma(\downarrow\uparrow)}{d\mtt} \right) \, ,
\label{difasud}
\end{equation}
where as above the first (second) arrow refers to the $t$ $({\bar t})$
spin projection onto  the reference axis
$ {\bf \hat a}$ $({\bf \hat b})$, and 
\begin{equation}
 N \equiv \frac{d\sigma(gg)}{d\mtt} + \frac{d\sigma(q {\bar
     q})}{d\mtt} \, .
\end{equation}
Here we take into account both the LO QCD and the weak contributions
to $N$. The relations (\ref{exaktres}) for the subprocesses $i$ imply that 
\begin{equation}
2 \Delta(\uparrow\downarrow) \, = \, N^{-1}\left(
  \frac{d\sigma(\uparrow, \, un ) - d\sigma(\downarrow, \, un )}{d\mtt}
- \frac{d\sigma(un,\, \uparrow) - d\sigma(un,\, \downarrow)}{d\mtt} \right) \, ,
\label{eq1dif}
\end{equation}
holds at the level of hadronic collisions. Here the first and second
(third and fourth)
term on the right-hand side of (\ref{eq1dif})
is the distribution of a $t \bar t$ sample
with $t$ $({\bar t})$ polarization parallel and anti-parallel 
to ${\bf \hat a}$ $({\bf \hat b})$
and  ${\bar t}$ ($t$) quarks with both spin projections added.
As already emphasized, Eq.~(\ref{eq1dif})
is simply a consequence of rotational invariance. \\
Now we choose the helicity axes (\ref{helbasis}). In this case we use the notation
\begin{equation}
Z_{RL} \,\equiv \,
 \frac{d\sigma_{+-}}{d\mtt} - 
\frac{d\sigma_{-+}}{d\mtt}  \, , \qquad \Delta_{RL} \equiv
\frac{Z_{RL}}{N} \, .
\label{difashel}
\end{equation}
Further we define the
$t$ and $\bar t$ single spin asymmetries in the helicity basis:
\begin{equation}
Z_{hel}\, \equiv \,  \frac{d\sigma_{+, \,un}}{d\mtt} - 
\frac{d\sigma_{-, \, un} }{d\mtt}  \, , \qquad
{\bar Z}_{hel}\, \equiv \,  \frac{d\sigma_{un,\, +}}{d\mtt} - \frac{d\sigma_{un, \, - }}{d\mtt}  \, ,
\label{sipsiashel1}
\end{equation}
and
\begin{equation}
\Delta_{hel}\, \equiv \, \frac{Z_{hel}}{N}\, , \qquad 
{\bar \Delta}_{hel} \, \equiv \, \frac{{\bar Z}_{hel}}{N}\     \, .
\label{sipsiashel2}
\end{equation}

Next we derive the consequences of CP invariance for these spin
observables. Let us first consider proton-antiproton collisions. 
If CP invariance holds then  (\ref{sspas}), (\ref{cp1rel}), (\ref {cpqgrel}), 
(\ref{eq1dif}), and the fact that $p \bar p$ is a CP eigenstate in its
c. m. frame imply  the relations
\begin{equation}
{\bar Z}_{hel}= - Z_{hel}\, , \qquad 
Z_{RL} = Z_{hel}= - {\bar Z}_{hel} \, , \qquad    \Delta_{RL} \, = \,
\Delta_{hel}\, = \, -  {\bar \Delta}_{hel}.
\label{cphad11}
\end{equation}
These relations hold also when CP-symmetric phase-space cuts are applied.
CP relations analogous to (\ref{cphad11}) can of course also be
derived for other, appropriately defined distributions. 
Eqs.~(\ref{cphad11}) hold to NLO
in the weak interactions. 

\begin{figure}[H]
\begin{center}
\epsfig{file=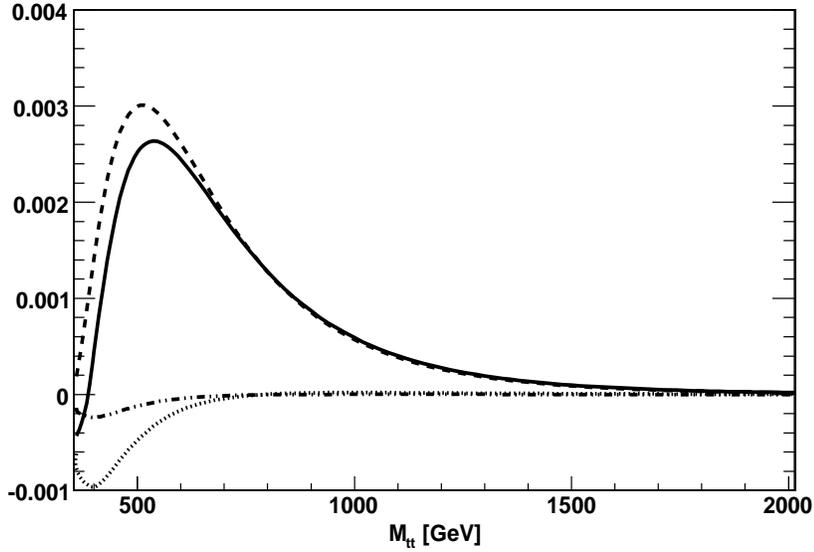, width=12cm,height=8cm }
\end{center}
\caption{The P-violating differential  spin 
asymmetry  $Z_{hel}$, 
defined in (\ref{sipsiashel1}), in units of [pb/GeV] 
at the LHC. Contribution from the $gg$ (dashed line) and $q \bar q$
(dotted line) subprocesses, and their sum (solid line). The
dash-dotted
line is the contribution from the $qg$ and ${\bar q}g$ subprocesses.
}\label{fig:ZRL}
 \end{figure}

\begin{figure}[H]
\begin{center}
\epsfig{file=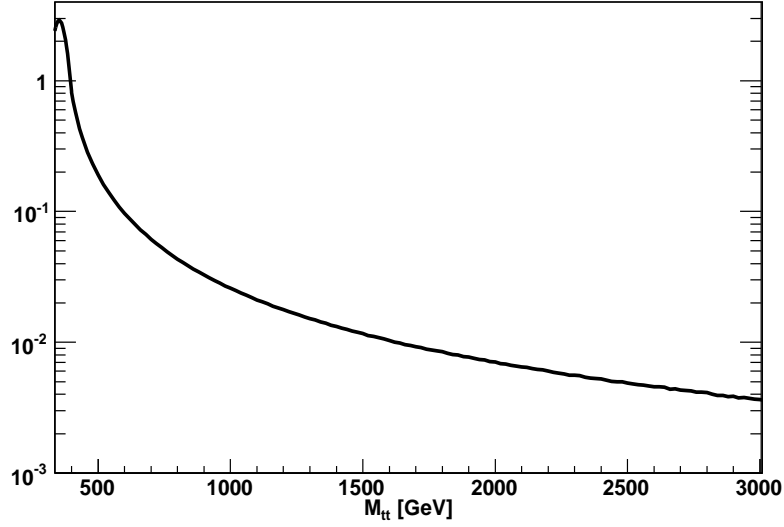, width=12cm,height=8cm }
\end{center}
\caption{The ratio $(-1)(Z_{hel}+{\bar Z}_{hel})/(Z_{hel}-{\bar
    Z}_{hel})$,  where $Z_{hel}$ and ${\bar Z}_{hel}$ are defined in
  (\ref{sipsiashel1}), at the LHC.}
\label{fig:qgZhel}
 \end{figure}

What about proton-proton collisions at the
LHC? As long as we take into account only $gg$ and $q \bar q$
initiated  $t \bar t$ production the relations
(\ref{cphad11})
are of course fulfilled. 
The single spin asymmetries (\ref{eq:qgbarspin})
 shown in  Fig.~\ref{fig:gqfh1} for top-quark pair production by
$Z$ boson exchange in $g q ({\bar q}) \to t {\bar t} q ({\bar q})$
lead to a violation of (\ref{cphad11}). This follows
from the result $\langle \langle 2 {\sp}\cdot {\bf\hat k} \rangle
{\rangle}_{qg} \neq \langle \langle 2 {\sp}\cdot {\bf\hat k} \rangle
{\rangle}_{{\bar q}g}$ and the CP relations (\ref{eq:DRLqgbarq}).
However, the parity-violating
contributions from these reactions 
to $Z_{hel}$ and ${\bar Z}_{hel}$ are, for large $\mtt$, small at the LHC, which we
shall show now.
Fig.~\ref{fig:ZRL} displays the contributions of the $gg$, 
$q \bar q$, $qg$ and ${\bar q} g$ subprocesses  to $Z_{hel}$ at the
LHC. We recall
that they are independent of the Higgs mass.
The virtual and real order $\alpha_s^2 \alpha$
corrections to the  $q \bar q$ subprocesses were  determined  in
 \cite{Bernreuther:2005is}  with  a phase-space slicing procedure.
The sum of these contributions to (\ref{sipsiashel1})
is infrared-finite and independent 
of the slicing parameter, as it should. In Fig.~\ref{fig:qgZhel}
the ratio $(-1)(Z_{hel}+{\bar Z}_{hel})/(Z_{hel}-{\bar Z}_{hel})$
is plotted as a function of $\mtt$. The numerator receives
contributions from the  $qg$ and ${\bar q} g$ subprocesses only,
while in the denominator all partonic subprocesses contribute.
Around $\mtt \simeq 400$ GeV this ratio is of order one, as the
contributions from $gg$ and $q \bar q$ tend to cancel each other, see
Fig.~\ref{fig:ZRL}. For larger $\mtt$ this ratio decreases rapidly in
magnitude. With this result we find that 
${\Delta}_{hel}+{\bar \Delta}_{hel}$ is about one per mill or less in
the whole $\mtt$ range; that is, the violation of the 
last relation of (\ref{cphad11}) at the LHC by SM weak interactions is very
small.

\begin{figure}[H]
\begin{center}
\epsfig{file=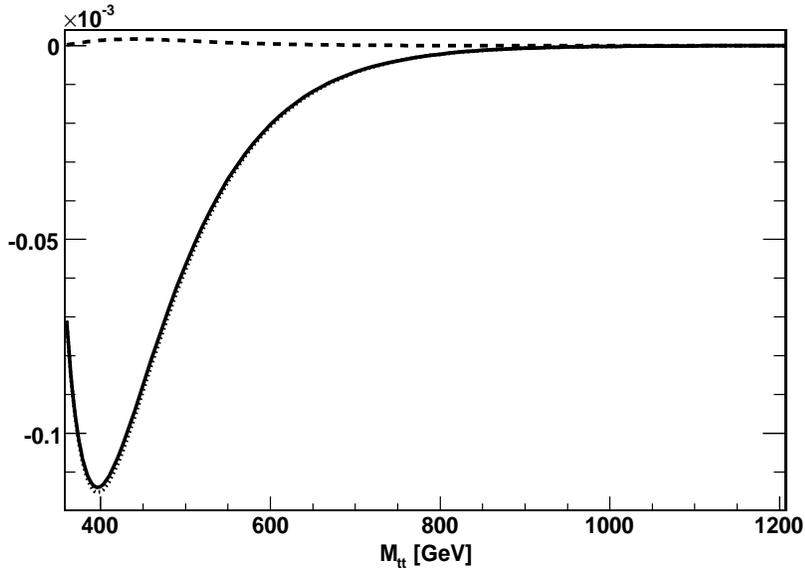, width=12cm,height=8cm }
\end{center}
\caption{The P-violating differential  spin 
asymmetries  $Z_{RL}=Z_{hel}$, 
defined in (\ref{difashel}), (\ref{sipsiashel1}), in units of [pb/GeV] 
at the Tevatron. Contribution from the $gg$ (dashed line) and $q \bar q$
(dotted line) subprocesses, and their sum (solid line). 
}\label{fig:ZRLtev}
 \end{figure}

\begin{figure}[H]
\begin{center}
\epsfig{file=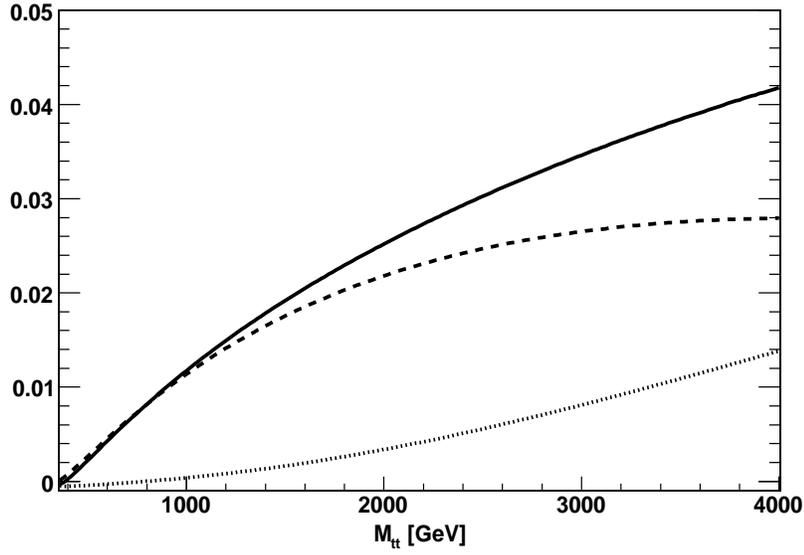, width=12cm,height=8cm }
\end{center}
\caption{The P-violating differential  spin 
asymmetry  $\Delta_{hel}$, defined in 
(\ref{sipsiashel2}), at the LHC. The dashed
and the dotted line is the contribution from the $gg$ and $q \bar q$
subprocesses, respectively, and the solid line is the sum of both terms. 
}\label{fig:Delrl}
 \end{figure}

\begin{figure}[H]
\begin{center}
\epsfig{file=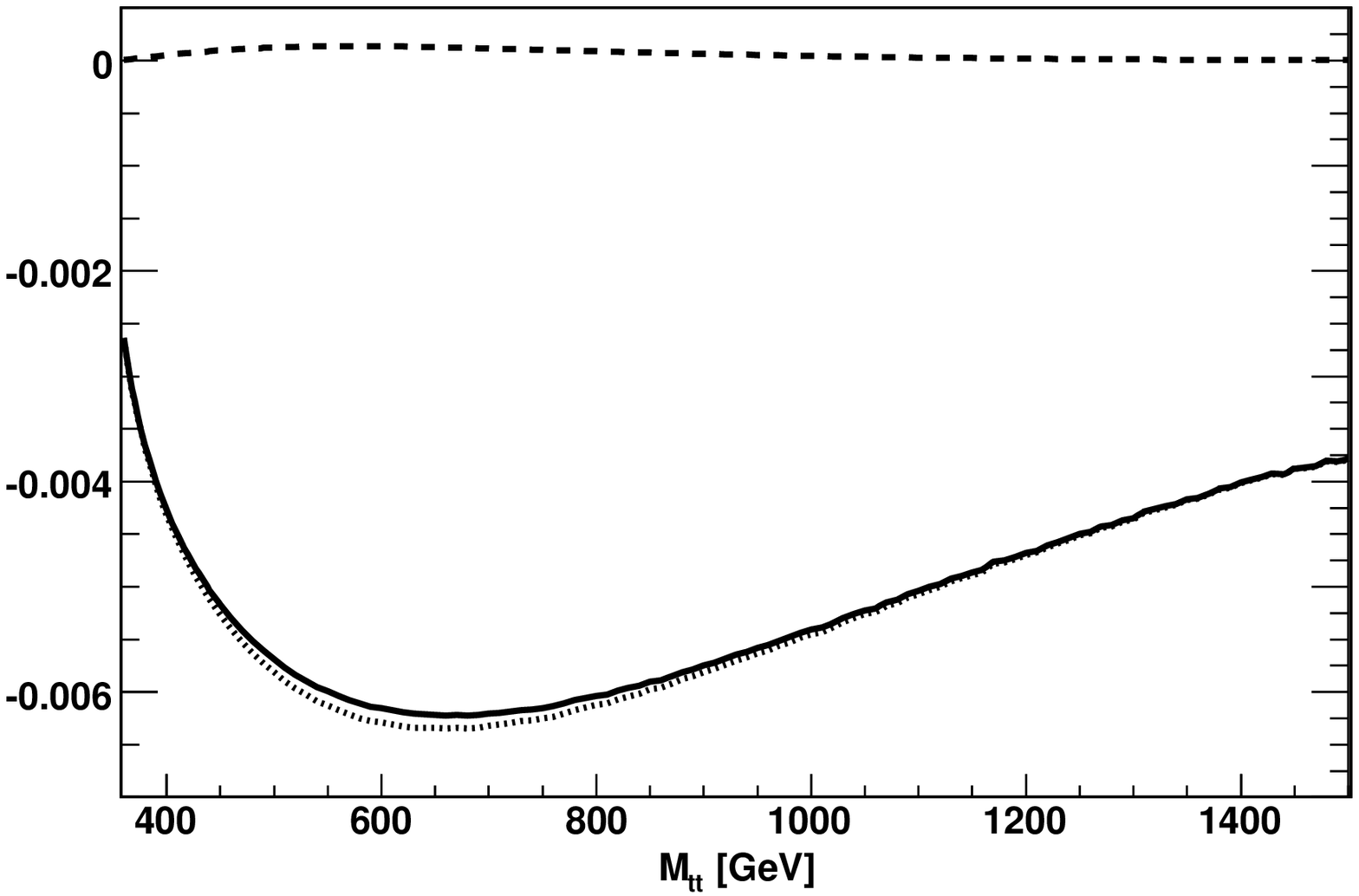, width=12cm,height=8cm}
\end{center}
\caption{The P-violating differential  spin 
asymmetries  $\Delta_{RL}=\Delta_{hel}$, defined in (\ref{difashel}),
(\ref{sipsiashel2}), at the Tevatron. The dashed
and the dotted line is the contribution from the $gg$ and $q \bar q$
subprocesses, respectively, and the solid line is the sum of both terms. 
}\label{fig:Delrltev}
\end{figure}

As the contribution of the $qg$ and ${\bar q}g$ 
subprocesses is small, it has been omitted in the next plots.
For the Tevatron  $Z_{hel}$ is shown in 
Fig.~\ref{fig:ZRLtev}. 
The ratio $\Delta_{hel}$ is
 displayed in  Fig.~\ref{fig:Delrl} and ~\ref{fig:Delrltev} for the
 LHC and the Tevatron, respectively. 
This asymmetry depends on the  Higgs mass  via the denominator $N$;
however, in the chosen range of $m_H$ this dependence is not visible
in the plots.  As discussed above, we have  $\Delta_{RL}=\Delta_{hel}$
for the Tevatron, and this relation holds also to very good
approximation for the LHC within the SM. That is, for the
LHC $\Delta_{RL}$ is given by the solid line in  Fig.~\ref{fig:Delrl}.

A parity-violating  double spin asymmetry proportional to $\Delta_{RL}$
was considered in detail first in  \cite{Kao:1999kj} for SM weak interactions 
to order $\alpha_s^2\alpha$, for a two-Higgs doublet model, and for the 
minimal supersymmetric extension of the SM. As far as SM weak interactions
are concerned, several contributions were
not taken into account in   \cite{Kao:1999kj}, namely, for $q \bar q$
annihilation, the infrared-divergent box contributions and the
corresponding real gluon radiation  and, for $gg$ fusion, the 
non-singlet neutral current contribution (which contributes to
$\sigma$, see Section 2).
The double-spin asymmetry $\delta {\cal A}_{LR}(\mtt)$ considered in that
paper and our $\Delta_{RL}$ are normalized differently, and the PDF
used in \cite{Kao:1999kj} are now outdated.
For these reasons a precise numerical  comparison of our results with those of
\cite{Kao:1999kj} is difficult. Let us compare the  results for
the integrated asymmetry 
\begin{equation}
A_{hel} \equiv \frac{ \int \left( d\sigma_{+,\, un} - d\sigma_{-,\, un}
    \right)}{\int d\sigma} 
\label{intarl}
\end{equation}
which is the integrated version of $\Delta_{hel}$ or
$\Delta_{RL}$, and which is equal
to the  quantity ${\cal A}$ of \cite{Kao:1999kj}.
For the LHC, using the cut $p_T>$ 100 GeV,
 the result of ref. \cite{Kao:1999kj} is
$|{\cal A}|=0.5\%$,  
while we obtain $A_{hel}=0.44 \%$. For the Tevatron, using
the  cut $p_T>$ 20  GeV, ref.  \cite{Kao:1999kj} obtained 
$|{\cal A}|=0.04\%$, 
while we get $A_{hel}=- 0.46 \%$.

The $gg$ contribution to the parity-violating  single
spin asymmetry in the helicity basis were computed for the LHC also 
in  \cite{Moretti:2006nf}. The quantity  ${\rm AL}_{tt}$
of that paper corresponds\footnote{The asymmetry
 ${\rm APV} \propto
d\sigma_{--} - d\sigma_{++}$ in Eq.~1 of \cite{Moretti:2006nf}  
is CP-odd and corresponds to 
$-\Delta_{CP}$ of Eq.~(\ref{cpas22}) below; i.e., its value
is zero to order $\alpha_s^2\alpha$ when taking only contributions
from $gg$ and $q \bar q$ subprocesses into account.}
to  $-Z_{hel}/[d\sigma(gg)/d\mtt]$.  We disagree with the
results shown in Figs.~2 and~3 of that paper.

Let us now discuss how these $t$ and $\bar t$ spin effects manifest
themselves at the level of the top-quark decay products. Of the main
$t \bar t$ decay modes, that is, the all-jets, lepton + jets, and dilepton
channels, probably only the latter two are useful for top-spin
physics, as the former has large backgrounds. The $t$, $\bar t$
polarizations and spin-spin correlations discussed above lead, through
the parity-violating weak decays of these quark, to characteristic
angular distributions and correlations among the final  state particles/jets.
According to the SM, in
semileptonic top-quark decays the outgoing charged lepton is the best top-spin
analyzer, while for non-leptonic top decays the resulting
least-energetic non-$b$ jet is a good and experimentally acceptable
choice \cite{Brandenburg:2002xr}. Thus, for measuring $t \bar t$ 
spin correlations at the
Tevatron or LHC one may consider the reactions
\begin{equation}
p {\bar p}, \; p p \to t {\bar t} \, + X \to a({\bf p}_+) + {\bar
  b}({\bf p}_-) \, + X\, ,
\label{abreact}
\end{equation}
where $a$ and ${\bar b}$ denotes either a charged lepton ($\ell=e,\mu)$ or
a jet from $t$ and $\bar t$ decay, respectively, and ${\bf p}_+$ and 
${\bf p}_-$ denote the 3-momenta of these particles/jets in the
respective $t$ and $\bar t$ rest frame\footnote{For the lepton + jets and
for the dileptonic channels the
$t$ and $\bar t$ momenta, i.e., their rest frames can be kinematically
reconstructed up to ambiguities which may be resolved with Monte Carlo
methods using the matrix element of the reaction.}. One may now 
choose two   polar vectors ${\bf{\hat a}}$ and  ${\bf{\hat b}}$
as reference axes, determine the angles 
$\theta_+ =\angle({\bf p}_+,{\bf{\hat a}})$ and
$\theta_- =\angle({\bf p}_-,{\bf{\hat b}})$ event by event, and 
consider the double distributions
\begin{equation}
\frac{1}{\sigma_{ab}}\frac{d\sigma}{d\cos\theta_+ d\cos\theta_-}
= \frac{1}{4}\left(1+ B_+\cos\theta_+ +  B_-\cos\theta_- -
C \cos\theta_+ \cos\theta_- \right) \, ,
\label{ddistab}
\end{equation}
where $\sigma_{ab}$ is the cross section of the channel (\ref{abreact}).
The right-hand side of (\ref{ddistab}) is the a priori form of this 
distribution if no cuts were applied. In the  presence of cuts the
shape of the distribution will in general 
be distorted; nevertheless, one may use
the bilinear form (\ref{ddistab}) as an estimator in fits to data.
The coefficient $C$ contains the information about the parity-even
$t \bar t$ spin correlations. 
These distributions were predicted for the Tevatron and the LHC 
in \cite{Bernreuther:2001rq,Bernreuther:2004jv} to  NLO
QCD for an number of reference axes, including the helicity axes
(\ref{helbasis}), in which case the corresponding $t \bar t$
spin correlation  is described by
${\cal O}_3$. It is straightforward to add to these NLO
QCD results the weak interaction corrections given by the 
function
$4\pi\alpha_s^2\alpha g_{gg}^{(1W,3)}/m_t^2$, defined in (\ref{eq:douevspin}) and shown in
Fig.~\ref{fig:r3gg}, using the formalism outlined in 
\cite{Bernreuther:2004jv}. This remark applies also
to the weak interaction corrections to ${\cal O}_4$, which induces
the opening angle distribution \cite{Bernreuther:2004jv} 
${\sigma_{ab}}^{-1} {d\sigma}/{d\cos\varphi}
= ( 1- D\cos\varphi)/2$, 
where $\varphi=\angle({\bf p}_+,{\bf p}_-)$. Adding up all 
$\mtt$ bins the effect of the weak interaction
corrections to $C_{hel}$ and to $D$ are not significantly larger than
the estimated experimental error of about $4\%$ at the LHC 
\cite{Hubaut:2005er}. As discussed above the weak interaction
contributions may be enhanced by suitable cuts on $\mtt$.

The information about the parity-odd, T-even top spin effects -- i.e.,
the single and double spin asymmetries (\ref{sspas}), 
(\ref{dupdown}) -- is contained in the coefficients $B_{\pm}$
of (\ref{ddistab}). The highest sensitivity to these effects
is achieved  with events
where the $t$ or $\bar t$ decay semileptonically. Consider the reactions
\begin{equation}
p {\bar p}, \; p p \to t {\bar t} \, +  X \to \ell^+({\bf p}_+) \, +   X\, ,
\label{ellreact}
\end{equation}
where $\ell=e, \mu$. Experimentally, the event
selection should discriminate against
single $t$ production, which also contributes to
the  final state (\ref{ellreact}). 
 Integrating (\ref{ddistab}) with respect to
$\cos \theta_-$ yields the distribution 
\begin{equation}
\frac{1}{\sigma_{\ell}}\frac{d\sigma}{d\cos\theta_+}
= \frac{1}{2}\left(1+ B_+\cos\theta_+ \right) \, .
\label{ddispvl}
\end{equation}
We consider here the helicity basis, which is
the best choice for the LHC. Thus 
$\theta_+ =\angle({\bf p}_+,{\bf{\hat k}})$, where
${\bf{\hat k}}$ is the $t$ quark direction in the $t \bar t$ ZMF.
For the computation of $B_+$ we need the unnormalized
decay density matrix $\rho$
for $t\to \ell^+ + X$, integrated over all  final-state
variables, except $\cos\theta_+$. It is given by
$ 2 \rho= \Gamma_\ell (I +\kappa_+ {\bf\sigma}\cdot {\bf\hat p}_+)$,
where $\sigma_i$ denote the Pauli matrices, 
$\Gamma_\ell$ is the semileptonic decay width and $\kappa_+$
is the top-spin analyzing power of $\ell^+$. In the SM
$\kappa_+=1$ to lowest order and $\kappa_+=0.9984$ including the order
$\alpha_s$ QCD corrections. With this
ingredient and with the results above we obtain
\begin{equation}
B_+ = \kappa_+ \frac{\int d\mtt\: Z_{hel}(\mtt)} {\sigma_{t \bar t}} \, ,
\label{predbpl}
\end{equation}
where $Z_{hel}$ is defined in (\ref{sipsiashel1})
 and $\sigma_{t \bar t}$ denotes the total $t\bar t$ cross section. 
With the results for $Z_{RL}= Z_{hel}$ displayed in
Figs.~\ref{fig:ZRL}  and Fig.~\ref{fig:ZRLtev}
we obtain the SM prediction for the parity-violating distribution 
(\ref{ddispvl}) for the LHC and the Tevatron
given in Table~\ref{tabapvns}. 
The distribution (\ref{ddispvl}) leads to  the
asymmetry
\begin{equation}
A_{PV} \equiv \frac{N_+ - N_-}{N_+ + N_-} = \frac{B_+}{2}
\label{asapv}
\end{equation}
where $N_\pm$ is the number of events (\ref{ellreact}) with
$\cos\theta_+$ larger or smaller than zero. If CP invariance holds,
$A_{PV}$ is equal to $A_{RL}$ defined in (\ref{intarl}).
The numbers for $A_{PV}$ for events at the Tevatron and at the
LHC with a $t \bar t$ invariant mass larger than $\mtt^*$ are given in Table~\ref{tabapvns}.
The statistical significance $S$ is estimated by $S\simeq
A_{PV} \sqrt{N_+ + N_-}$, where the number of dileptonic $\ell^+
\ell'^-$ ($\ell =e,\mu$, $\ell'=e, \mu, \tau$) and
$\ell^+$ + jets events, which constitute  a fraction of about $2/9$ of
all $t\bar t$ events. It has been computed assuming an integrated
luminosity of 10 (fb)$^{-1}$ and  100 (fb)$^{-1}$ for the Tevatron and
the LHC, respectively. For the LHC one obtains $S>4$ for suitable
cuts. It remains to be investigated
with which precision  $A_{PV}$ can actually be
measured by an LHC experiment. 
\begin{table}[ht]
\caption{Standard Model prediction for the parity-violating asymmetry
  (\ref{asapv}) for the Tevatron and the LHC, and 
the statistical significance $S$.} \label{tabapvns} 
\bc
\begin{tabular}{|c|c|c|c|c|c|}
\hline
\hhline{------}
 $ M_{t\overline{t}}^*  \:\left[{\rm GeV}\right]$    & $A_{PV}$, Tevatron & S& $
 M_{t\overline{t}}^*\: \left[{\rm GeV}\right]$ & $A_{PV}$, LHC  &S\\
\hline
 400              &    $-0.0027$     &  0.2&   500        &  0.0028 & 5.4\\
\hline
 700              &      $-0.0030$    &  0.04  &  1000        & 0.0077 & 3.7 \\
\hline 
1000              &      $-0.0026$    &  0.006  &  1500        & 0.011
& 1.9\\
\hline
\end{tabular}
\ec
\end{table}

If one uses events where the $\bar t$ quarks have
decayed  semileptonically,
\begin{equation}
p {\bar p}, \; p p \to t {\bar t} \, +  X \to \ell^-({\bf p}_-) \, +   X\, ,
\label{aellreact}
\end{equation}
the analogue of  (\ref{ddispvl}) is 
\begin{equation}
\frac{1}{\sigma_{\ell}}\frac{d\sigma}{d\cos\theta_-}
= \frac{1}{2}\left(1+ B_-\cos\theta_- \right) \, ,
\label{addispvl}
\end{equation}
where $\theta_- =\angle({\bf p}_-,{\bf \hat k}_{\bar t})$, and
it is to be recalled that  ${\bf \hat k}_{\bar t}
=- {\bf \hat k}$ in the $t \bar t$ ZMF. CP invariance, 
which holds to the order of
perturbation theory employed here, implies that the $\bar t$ decay
density matrix ${\bar\rho}({\bar t}\to \ell^-)$ is of the form 
$ 2 {\bar \rho}= \Gamma_\ell (I - \kappa_+ {\bf\sigma}\cdot {\bf\hat
  p}_-)$. Using this and (\ref{cphad11}) we obtain
\begin{equation}
B_- = B_+ \, .
\label{bpmcprel}
\end{equation}
The Standard Model predictions of
Figs.~\ref{fig:ZRL},~\ref{fig:ZRLtev} 
and of Table~\ref{tabapvns} may be used
as reference values in future searches for parity-violating effects in
hadronic $t \bar t$ production and decay. 
Apart from new physics effects in $t \bar t$ production, also 
non-SM effects in top decay may influence the 
distributions (\ref{ddispvl}), (\ref{addispvl}).  As the
charged-lepton coefficient $\kappa_+$ is maximal in the SM,
it may be decreased by new interactons. For instance, if $t\to b \ell^+
\nu_{\ell}$ would be solely mediated by the exchange of a charged Higgs 
boson then $\kappa_b =1$ and $\kappa_+ < 1$, which would lead to a
smaller 
$A_{PV}$. Thus larger values of $A_{PV}$ than those given in 
Table~\ref{tabapvns} would point towards non-SM parity violation in $t
\bar t$ production. Effects larger than those given in
Table~\ref{tabapvns}
are possible for instance in two-Higgs doublet or supersymmetric
models if the new particles are not too heavy \cite{Kao:1999kj}.
As to new physics effects in  polarized semileptonic top decay
mediated by $W$ exchange, one
should note the following: if these new interactions lead only to anomalous
form factors in the $tWb$ vertex, this would
not change the lepton angular distribution 
\cite{Grzadkowski:1999iq,Rindani:2000jg}, i.e. $\kappa_+$,
as long as these anomalous form factors are small. On the other hand the
energy distribution $d\Gamma/dE_\ell$, which we did not use in our analysis,
may change. For supersymmetric QCD corrections to the $tWb$ vertex the deviations
from the SM are, however, negligible \cite{Brandenburg:2002xa}.

For completeness,  we briefly discuss  a  
differential distribution which results from  the  $gg$ and $q {\bar q}$
CP asymmetries (\ref{eq:cpas}). It reads in the helicity basis:
\begin{equation}
\Delta_{CP} \,\equiv \,
N^{-1}\left( \frac{d\sigma_{++}}{d\mtt} - 
\frac{d\sigma_{--}}{d\mtt} \right) \, = \,  \frac{1}{2} \left( \Delta_{hel} +  {\bar
\Delta}_{hel} \right)\, .
\label{cpas22}
\end{equation}
The second equality is due to rotational invariance. As already
emphasized this asymmetry is a probe of non-standard CP violation in
$t \bar t$ production. A non-zero value of $\Delta_{CP}$ is equivalent
to a violation of  (\ref{cphad11}). In order to check for CP violation
with this variable at the level of the top-quark decay products,
one strategy would be to
compare the distributions (\ref{ddispvl}) and (\ref{addispvl}) for
an event sample (\ref{ellreact}) and  (\ref{aellreact}), respectively,
i.e., to check for a violation of (\ref{bpmcprel}). 
If, for instance, non-standard heavy neutral Higgs
boson(s) $\varphi$ with mass $m_{\varphi}\gtrsim 2 m_t$ and with
scalar and pseudoscalar Yukawa couplings to top quarks exist, 
$\Delta_{CP}$  and likewise $B_+ -B_-$ can be of the order of several percent in magnitude
around $\mtt \sim m_{\varphi}$ as was shown in \cite{BBF4}.
As discussed above, at the LHC there are SM contributions to (\ref{cpas22}) 
from the $q g$ and ${\bar q} g$ subprocesses, but this amounts to less than
one per mill.

\section{Conclusions}
The main interest in the SM weak interaction corrections to hadronic $t
\bar t$ production is the determination of 
their  size at large transverse top-momentum and/or large
$t \bar t$ invariant mass (i.e., the weak Sudakov effects)  and of the
parity-violating effects, especially at the LHC.  
In this paper we have calculated the one-loop weak corrections 
to top quark pair
production due to gluon-gluon fusion and (anti)quark-gluon
scattering.  This gives, 
together with our previous result  for
$q {\bar q} \to t \bar t (g)$ 
\cite{Bernreuther:2005is}, the complete 
corrections of order $\alpha_s^2\alpha$ to  $t \bar t$
production with $t$ and $\bar t$
polarizations  and spin-correlations fully taken into account. 
For $t \bar t$ production at the Tevatron and at LHC we have
determined  the weak
contributions  to the transverse top-momentum and to the 
$t \bar t$ invariant mass distributions.  For the LHC the size of the
weak corrections to $d\sigma/dp_T$ and $d\sigma/d\mtt$ is of the order of
10 percent for large $p_T$ and $\mtt$, respectively, as compared with LO results. 
Further we have computed the
order $\alpha_s^2\alpha$ contributions to two parity-even $t \bar t$ spin
correlation observables which are of interest for the LHC.
As far as parity-violating effects are concerned we derived,
for CP-invariant interactions, relations between 
 parity-violating double and single top spin asymmetries.
We pointed out how one may probe in this context for non-standard CP
 violation, and we computed the SM background to
an appropriate observable for the LHC. The 
parity-violating effects are best  analyzed for $t
 \bar t$ events where the $t$ $(\bar t)$ quark decays
 semileptonically, and we have computed a charged-lepton forward-backward asymmetry $A_{PV}$
with respect to the $t$ $(\bar t)$ quark direction. At the LHC
 $A_{PV}$ is of the order of one percent for suitable cuts on 
$\mtt$. Whether such a small asymmetry can be measured at the LHC
remains to be investigated by experimentalists with simulations including detector effects.
Nevertheless, this result should serve, like the predictions for
$d\sigma_{weak}/dp_T$ and $d\sigma_{weak}/d\mtt$,  as a reference in the detailed
exploration of the top quark interactions with future data.

\subsubsection*{Acknowledgements}
We thank  A. Scharf and P. Uwer for discussions and 
for communication of their results
prior to publication.  Further we thank S. Moretti
and D. Ross for a correspondence
 concerning \cite{Moretti:2006nf}. W.B. wishes to thank also the Physics 
Department of Shandong University, Jinan, where part of this
work was done, for its hospitality. This work was supported
by Deutsche Forschungsgemeinschaft (DFG) SFB/TR9, by
DFG-Graduiertenkolleg RWTH Aachen, by NSFC, by NCET, and by 
Huoyingdong Foundation, China.

\newpage
%
%
%
%
%
%
%
%
%
%
%
%

%

%
%
%
%
%
%
%
%
%
%
%
%
%
%

\end{document}